\documentclass[aps,a4paper,preprint,amsfonts]{revtex4}
\usepackage[]{bm,epsfig}
\usepackage[T1]{fontenc}
\newcommand{\clc}[1]{\multicolumn{1}{c}{#1}}

\begin{document}

\title{Interaction between LiH molecule and Li atom from
       state-of-the-art electronic structure calculations}

\author{\sc Wojciech Skomorowski}
\author{\sc Filip Paw\l owski}
\author{\sc Tatiana Korona}
\author{\sc Robert Moszynski\footnote[1]{Author for correspondence;
e-mail:robert.moszynski@tiger.chem.uw.edu.pl}}
\affiliation{\sl Faculty of Chemistry, University of Warsaw, Pasteura 1,
02-093 Warsaw, Poland}

\author{\sc Piotr S. {\.Z}uchowski}
\author{\sc Jeremy M. Hutson}
\affiliation{\sl Department of Chemistry, Durham University,
South Road, Durham DH1 3LE, United Kingdom}
\begin{abstract}
State-of-the-art {\em ab initio} techniques have been applied
to compute the potential energy surface for the lithium atom
interacting with the lithium hydride molecule in the
Born-Oppenheimer approximation. The interaction potential was
obtained using a combination of the explicitly correlated
unrestricted coupled-cluster method with single, double, and
noniterative triple excitations [UCCSD(T)--F12] for the
core-core and core-valence correlation and full configuration
interaction for the valence-valence correlation. The potential
energy surface has a global minimum 8743 cm$^{-1}$ deep if the
Li--H bond length is held fixed at the monomer equilibrium
distance or 8825 cm$^{-1}$ deep if it is allowed to vary. In
order to evaluate the performance of the conventional CCSD(T)
approach, calculations were carried out using
correlation-consistent polarized valence $X$-tuple-zeta basis
sets, with $X$ ranging from 2 to 5, and a very large set of
bond functions. Using simple two-point extrapolations based on
the single-power laws $X^{-2}$ and $X^{-3}$ for the orbital
basis sets, we were able to reproduce the CCSD(T)--F12 results
for the characteristic points of the potential  with an error
of 0.49\% at worst. The contribution beyond the CCSD(T)--F12
model, obtained from full configuration interaction (FCI)
calculations for the valence-valence correlation, was shown to
be very small, and the error bars on the potential were
estimated. At linear LiH--Li geometries the ground-state
potential shows an avoided crossing with an ion-pair potential.
The energy difference between the ground-state and
excited-state potentials at the avoided crossing is only 94
cm$^{-1}$. Using both adiabatic and diabatic pictures we
analyse the interaction between the two potential energy
surfaces and its possible impact on the collisional dynamics.
%An analysis of the nonadiabatic
%coupling matrix elements suggests that the interaction with the excited
%state should have little its potential impact on the collisional dynamics
%, because
%the couplings are important only at the linear geometry. Further
%analysis of the diabatic potentials confirms this analysis.
When the LiH bond is allowed to vary, a seam of conical intersections appears at
$C_{\rm 2v}$ geometries. At the linear LiH--Li geometry, the conical
intersection is at a Li--H distance which is only slightly larger than
the monomer equilibrium distance, but for nonlinear geometries it
quickly shifts to Li--H distances that are well outside the classical
turning points of the ground-state potential of LiH. This suggests that
the conical intersection will have little impact on the dynamics of
Li--LiH collisions at ultralow temperatures. Finally, the reaction
channels for the exchange and insertion reactions are also analyzed,
and found to be unimportant for the dynamics.
\end{abstract}

\maketitle

\section{Introduction}
\label{sec1} Ultracold molecules offer new opportunities for scientific
exploration, including studies of molecular Bose-Einstein condensates,
novel quantum phases, and ultracold chemistry. For molecular
interactions that take place at microKelvin temperatures, even the
smallest activation energy exceeds the available thermal energy. This
opens up new possibilities for controlling the pathways of chemical
reactions (see, e.g., Ref.\ \cite{Krems:08}).

A major objective of current experiments on cold molecules is to
achieve quantum degeneracy, particularly for polar molecules. Two
approaches are being pursued: indirect methods, in which molecules are
formed from pre-cooled atomic gases, and direct methods, in which
molecules are cooled from room temperature. There have been very
substantial recent advances, particularly in indirect methods. In
particular, the JILA \cite{Ni:KRb:2008} and Innsbruck
\cite{Danzl:ground:2010} groups have formed deeply bound ground-state
molecules at temperatures below 1 $\mu$K, by magnetoassociation of
pairs of ultracold atoms followed by coherent state transfer with
lasers. Methods that form ultracold molecules from ultracold atoms are
however restricted at present to species formed from heavy alkali-metal
atoms.

Direct methods, such as buffer-gas cooling \cite{Doyle:98}, Stark
deceleration \cite{Meijer:99}, crossed-beam collisional cooling
\cite{Elioff} and Maxwell extraction \cite{Buuren}, are applicable to a
much larger variety of chemically interesting molecules. However, these
methods cannot yet reach temperatures below 10 to 100~mK. Finding a way
to cool these molecules further, below 1~mK, is one of the biggest
challenges facing the field. The most promising possibility is
so-called sympathetic cooling, in which cold molecules are introduced
into an ultracold atomic gas and thermalize with it. Sympathetic
cooling has been successfully used to achieve Fermi degeneracy in
$^6$Li \cite{DeMarco:1999} and Bose-Einstein condensation in $^{41}$K
\cite{Modugno:2001}, and for producing ultracold ions
\cite{Zipkes2010,Zipkes2010a,Schmid}. However, it has not yet been
achieved for molecular systems, although there are theoretical
proposals for experiments in which ultracold NH or ND$_3$ molecules are
obtained by collisions with a bath of colder atoms such as Rb, Mg or N
\cite{Zuchowski:NH3:2009,Soldan:MgNH:2009,Wallis:MgNH:2009}.

The group at Imperial College London recently succeeded in producing
samples of cold LiH molecules in the first rotationally excited state
\cite{Tokunaga:2007,Tokunaga:2009} using Stark deceleration. LiH is an
attractive molecule for cooling, since it has large dipole moment and
light mass, so that it can be controlled easily with fields. It has a
relatively large rotational constant (7.5 cm$^{-1}$), which opens up
the possibility of producing cold molecules in a single excited
rotational state. There is proposal to produce ultracold LiH molecules
by sympathetic cooling with Li \cite{Tarbutt:privatecomm}. However,
sympathetic cooling can be successful only if the rate of elastic
(thermalization) collisions is large compared to the rate of inelastic
(deexcitation) collisions, which cause trap loss. The main objects of
the present paper are to explore the interaction between Li atoms and
LiH molecules, to understand the nature of the interaction between
these two species, and to obtain a detailed and accurate potential
energy surface for the Li--LiH system.

The results of scattering calculations at ultralow temperature are very
sensitive to the details of the interaction potential
\cite{Zuchowski:NH3:2009,Wallis:MgNH:2009}. For systems containing
heavy atoms, the methods of quantum chemistry currently available
cannot generate interaction potentials with accuracy better than a few
percent. This limitation is caused by approximate treatments of
correlation effects and relativistic contributions. With potential
energy surfaces of moderate precision, it is usually possible to
extract only qualitative information from low-energy collision
calculations. By contrast, Li--LiH is a light system containing only 7
electrons and state-of-the-art {\em ab initio} electronic structure
calculations can be performed with no significant approximations. It
therefore offers a unique possibility to produce a very precise
interaction potential, which will allow a quantitative description of
Li--LiH collision dynamics, even in the ultralow temperature regime.

In electronic structure calculations one aims at approaching
the exact solution of the Schr{\"o}dinger equation, as closely
as possible within the algebraic approximation. In practice,
this is accomplished by combining hierarchies of one-electron
and $N$-electron expansions. The accuracy increases across the
hierarchies in a systematic manner, allowing the errors in the
calculations to be controlled and a systematic approach to the
the exact solution to be achieved. The standard $N$-electron
hierarchy employed in electronic structure calculations
consists of the Hartree-Fock (HF), second-order M\o
ller-Plesset perturbation theory (MP2), coupled-cluster with
single and double excitations (CCSD), and coupled-cluster with
single, double, and approximate noniterative triple excitations
[CCSD(T)] models, with the latter recovering most of the
correlation energy. Thus, CCSD(T) constitutes a robust and
accurate computational tool nowadays. All these models are
size-consistent, which means that the interaction potential
shows the correct dissociation behaviour at large
intermolecular distances. In contrast, methods based on the
configuration interaction approach with a restricted excitation
space like multireference configuration interaction limited to
single and double excitations (MRCISD) are not size-consistent
and therefore they are not well suited for calculations of the
interaction energy.
%The CCSD method recovers most of the correlation energy and constitutes
%a robust and accurate computational tool. The CCSD(T) model adds a
%noniterative correction for triple excitations to the CCSD energy,
%bringing the accuracy to yet higher level. The CCSD(T) approach has
%been shown to yield outstandingly good-quality reaction enthalpies
%\cite{atomization_energies_and_enthalpies}. The scaling of
%computational time with the size of the molecule ($M$), increases
%across the hierarchy HF, MP2, CCSD, CCSD(T) as $M^4$, $M^5$, $M^6$, and
%$M^7$, respectively.
%Going beyond this hierarchy, towards models with even higher accuracy,
%raises the computational cost even further. In particular,
%computational cost prevents the routine application of the full
%configuration interaction (FCI) model (which yields an exact solution
%for a given one-electron basis set). Nevertheless, the FCI model can be
%applied to small systems, provided the one-electron basis set is not
%too large.

The most popular example of a one-electron hierarchy is the family of
Dunning correlation-consistent polarized valence basis sets, cc-pV$X$Z
\cite{Dunning:89} with the cardinal number $X$ going from D (double-zeta), through
T indicating triple-zeta, and so on.
These have successfully been combined with the HF,
MP2, CCSD, CCSD(T) hierarchy of wave function models for the
calculation of various molecular properties \cite{electronic_energies,basis,static_dipole_moments}.
% and other
%properties; see, for example, Refs.\ \cite{basis} and
%\cite{static_dipole_moments}.
%Increasing the cardinal number $X$
%improves the quality of the basis set: $X=$ D indicates double-zeta
%quality, $X=$ T indicates triple-zeta, and so on. To reach the
%basis-set limit, and thus to make the one-electron expansion complete,
%would require taking $X$ to infinity, which is clearly impractical.
%Nevertheless,
The basis-set limit, corresponding to $X\to\infty$, may be approached either by
extrapolating the results obtained with finite cardinal numbers towards
infinite $X$ \cite{extrap1,extrap2}, or by replacing the standard
one-electron hierarchy by explicitly-correlated methods, such as
CCSD--F12 and CCSD(T)--F12
\cite{skhv08a,skhv08b,Koehn:2008,tknh07,bokhan2008,tkh2008}, in which
the interelectron distance $r_{12}$ is explicitly introduced into the
wave function\cite{Kutzelnigg:1985,Klopper:1987,Ten-no:2004}.
The F12 methods have recently been implemented
efficiently \cite{efficient_ccsd_f12,ccf12,F12Rev,F12Rev_Tew} and shown
to accelerate the convergence towards the basis-set limit for a number
of properties
\cite{response_properties_from_f12,nh07,geometries_and_frequencies_from_f12}.

In the present paper, we combine all-electron spin-unrestricted CCSD(T)--F12 calculations
with frozen-core FCI calculations to yield a highly accurate best
estimate of the Li--LiH interaction potential. We also compare the F12 interaction energies with
results obtained from standard (not explicitly correlated) CCSD(T) calculations.
We then characterize the
ground-state potential, analyze possible interactions with excited
states, and investigate channels for reactive collisions.

\section{Computational Details}
\label{sec2} We have calculated the interaction energies between the
lithium atom and the lithium hydride molecule in Jacobi coordinates
($R,r,\theta$), defined for the isotopic combination
$^7$Li--$^7$Li$^1$H. Calculations were performed for states of
$^2A^\prime$ symmetry in the $C_{s}$ point group. The LiH bond
distance, $r$, was initially kept frozen at the LiH monomer equilibrium
distance of 3.014 bohr \cite{lihre}. The distance $R$ between Li and
the center of mass of LiH ranged from 3.0 to 10.0 bohr with an interval
of 0.5 bohr, and then from 11.0 to 20.0 bohr with an interval of 1.0
bohr. Additional distances of 30.0, 40.0, and 50.0 bohr were also used.
The angle $\theta$, between the vector pointing from Li to H in the LiH
molecule and the vector pointing from the center of mass of the
molecule to the Li atom, was varied from $0^{\circ}$ to $180^{\circ}$
with an interval of $15^{\circ}$; $\theta=0^{\circ}$ corresponds to
Li--H---Li configurations. We thus used a total of 28 intermonomer
distances, $R$, which combined with the 13 values of $\theta$ yielded
364 grid points on the two-dimensional interaction energy surface.

Calculations with uncorrelated basis functions were carried out using the unrestricted version of the
coupled-cluster model CCSD(T)
with Dunning's cc-pV$X$Z(-mid) basis sets with $X=$ D, T, Q, 5, where
mid indicates the inclusion of an additional set of basis functions,
the so-called midbond-95 set \cite{mid_partridge},
placed at the middle of the Li--LiH distance $R$. All electrons were
correlated in these calculations. Additionally, for the purpose of
comparison with the FCI results (see below), the frozen-core
approximation ($1\sigma_{\mathrm{LiH}}$ and $1s_{\mathrm{Li}}$ orbitals
kept frozen) was used for the cc-pVQZ basis. All these calculations
were carried out using the {\sc molpro} package \cite{MOLPRO2008}.
The full basis set of the dimer
was used in the supermolecular calculations and the Boys and Bernardi
scheme \cite{Boys:70} was used to correct for basis-set superposition
error.

The explicitly correlated spin-unrestricted CCSD--F12 and CCSD(T)--F12
\cite{ccf12,kw08,efficient_ccsd_f12,kaw2009}
calculations were carried out with the {\sc molpro} code
\cite{MOLPRO2008} to establish the CCSD and CCSD(T) basis-set limits
for the LiH--Li interaction.
We chose to use the F12b variant \cite{ccf12,kaw2009} of the explicitly
correlated spin-unrestricted energy
implemented in the {\sc molpro} code.
Employing the fixed-amplitude ansatz for
the F12 wave function ensured the orbital invariance and
size-consistency of the CCSD-F12 and CCSD(T)-F12 results. The QZVPP
basis set \cite{def2} was employed as the orbital basis in the F12
calculations. The corresponding QZVPP-jk basis set \cite{jk} was used
as the auxiliary basis for the density-fitting approximation
\cite{df1,kw08} for many-electron integrals, while the uncontracted
version of the QZVPP-jk basis was used to approximate the
Resolution-of-Identity in the F12 integrals \cite{ks2002,v2004}. In
addition, the valence correlation in the dimer was described with the
full configuration interaction method (FCI). The FCI and standard CCSD(T) calculations
in the frozen-core approximation were carried out using the cc-pVQZ basis.
The {\sc dalton} package \cite{dalton20} and the {\sc
lucia} program \cite{lucia} were combined to yield the FCI results.

To calculate potential energy surface $V(R,\theta)$ with the LiH bond length kept fixed
at its equilibrium value
we used computational scheme which was previously applied in theoretical studies of
the ground and excited states of the calcium dimer \cite{buss0,buss1,buss2,buss3,Koch:08}.
The potential $V(R,\theta)$ was constructed according to the following expression:
\begin{equation}
V(R,\theta) =
V^{\rm CCSD(T)-F12}(R,\theta) + \delta V_{\rm v-v}^{\rm FCI}(R,\theta)
\label{Pig}
\end{equation}
where  $V^{\rm CCSD(T)-F12}(R,\theta)$ contribution was
obtained from all-electron CCSD(T)-F12 calculations, while the
correction for the valence-valance correlation beyond the
CCSD(T)-F12 level,  $\delta V_{\rm v-v}^{\rm FCI}(R,\theta)$,
was calculated in an orbital cc-pVQZ basis set. Both terms,
$V^{\rm CCSD(T)-F12}(R,\theta)$ and $\delta V_{\rm v-v}^{\rm
FCI}(R,\theta)$, were obtained from the standard expressions
for the supermolecule interaction energy, as given in Ref.
\cite{buss3}.

The long-range asymptotic form of the potentials is of primary
importance for cold collisions. We have therefore computed the leading
long-range coefficients that describe the induction and dispersion
interactions up to and including $R^{-10}$ and $l=4$ terms,
\begin{equation} V(R,\theta)= -\sum_{n=6}^{10}\sum_{l=0}^{n-4}
\frac{C_n^l}{R^n}P_l(\cos\theta),
\label{lr1}
\end{equation} where $l$ is even/odd for $n$ even/odd, and
$C_n^l=C_n^l({\rm ind})+C_n^l({\rm disp})$. The long-range coefficients
$C_n^l({\rm ind})$ and $C_n^l({\rm disp})$ are given by the standard
expressions (see, e.g., Refs.\ \cite{Jeziorski:94,Moszynski:08}). The
multipole moments and polarizabilities of LiH were computed with the
recently introduced explicitly connected representation of the
expectation value and polarization propagator within the
coupled-cluster method \cite{Jeziorski:93,Moszynski:05,Korona:06a},
while the Li polarizabilities (both static and at imaginary
frequencies) were taken from highly accurate relativistic calculations
from Derevianko and coworkers \cite{Derevianko:10}.

The interaction potentials were interpolated between calculated points
using the reproducing kernel Hilbert space method (RKHS) \cite{rkhs}
with the asymptotics fixed using the {\em ab initio} long-range Van der
Waals coefficients. The switching function of Ref.\ \cite{Janssen} was
used to join the RKHS interpolation smoothly with the Van der Waals
part in the interval between $R_a=18$ and $R_b=26$ bohr.

\section{Convergence of the Li--LiH interaction potential
towards the exact solution}

In sec.~\ref{sub1} we analyze the convergence of the Li--LiH
interaction potential with respect to the one-electron and $N$-electron
hierarchies. Based on the analysis, we give in sec.~\ref{sub2} our best
estimate for the ground-state interaction potential with the Li--H bond
length fixed at its monomer equilibrium value. The features of the
potential are presented in sec.~\ref{sub3}.

\subsection{Convergence of the one-electron and $N$-electron hierarchies}
\label{sub1}

In order to investigate the saturation of the Li--LiH interaction
energy in the one-electron space, we have analyzed three characteristic
points of the Li--LiH potential (the global minimum, the saddle point,
the local minimum,  and one point very close to  
the avoided crossing: $R=5.5$ bohr and $\theta=0.0^\circ$). 
The characteristic points were obtained from
the potentials
calculated at the CCSD(T) / cc-pV$X$Z-mid level of theory, for $X =$ D,
T, Q, and 5. The interaction energies were then compared to the
corresponding energies of the spin-unrestricted CCSD(T)-F12 / QZVPP potential
(approximation F12b), which
serves as the basis-set limit. To evaluate the accuracy of the pure
one-electron basis ({\em not}\/ explicitly correlated), the relative
percentage errors, $\Delta_{\rm F12b} = (V^{{\rm cc-pV}X{\rm
Z}} - V^{\rm F12b}) \, / \, |V^{\rm F12b}| \, \cdot
\, 100\%$, were determined for each $X$ at every characteristic point.
The results are given in Table \ref{tab0}.

We have also evaluated the characteristic points from the extrapolated
interaction energy surfaces, which were generated as follows: at each
grid point, the extrapolated {\em total}\/ energies for Li, LiH, and
Li--LiH were obtained by adding the Hartree-Fock energy calculated with
cardinal number $X$ to the extrapolated correlation energy, $E^{\rm
corr}_{(X-1)X}$, obtained from the two-point extrapolation formula
\cite{extrap1,extrap2},
\begin{equation}
\label{extrapol}
E^{\rm corr}_{(X-1)X} = E^{\rm corr}_X + \frac{E^{\rm corr}_X
- E^{\rm corr}_{(X-1)}}{[1-(X)^{-1}]^{-\alpha}-1},
\end{equation}
where $E^{\rm corr}_{(X-1)}$ and $E^{\rm corr}_X$ are the correlation
energies obtained for two consecutive cardinal numbers, $(X-1)$ and
$X$, respectively. The final extrapolated interaction energy at a
single grid point is obtained by subtracting the Li and LiH
extrapolated total energies from the Li--LiH extrapolated total energy.
We used the values $\alpha = 2$ and $\alpha = 3$, which were
recommended by Jeziorska {\em et al.}\/ in their helium dimer study
\cite{mj1,mj2} as the ones most suited for extrapolating all the
components of the interaction energy. The energies of the
characteristic points obtained in this way were compared with the
CCSD(T)-F12 / QZVPP results and the corresponding values of
$\Delta_{\rm F12b}$ are included in Table \ref{tab0}.

\begin{figure}
\vspace*{2cm}
\begin{center}
\begin{tabular}{ccc}
{\includegraphics[scale=0.30,angle=0]{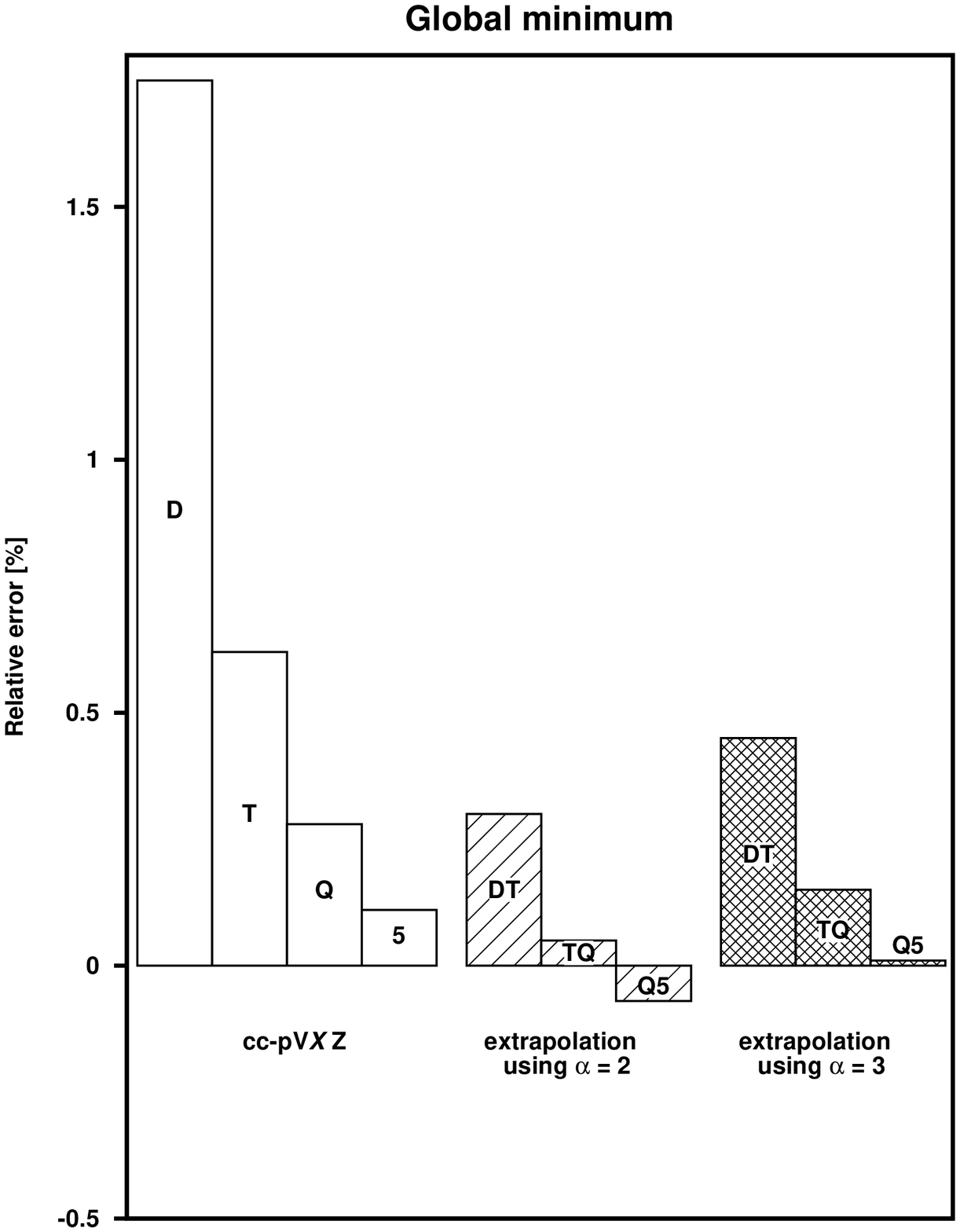}} &\hspace*{1.0cm}
{\includegraphics[scale=0.30,angle=0]{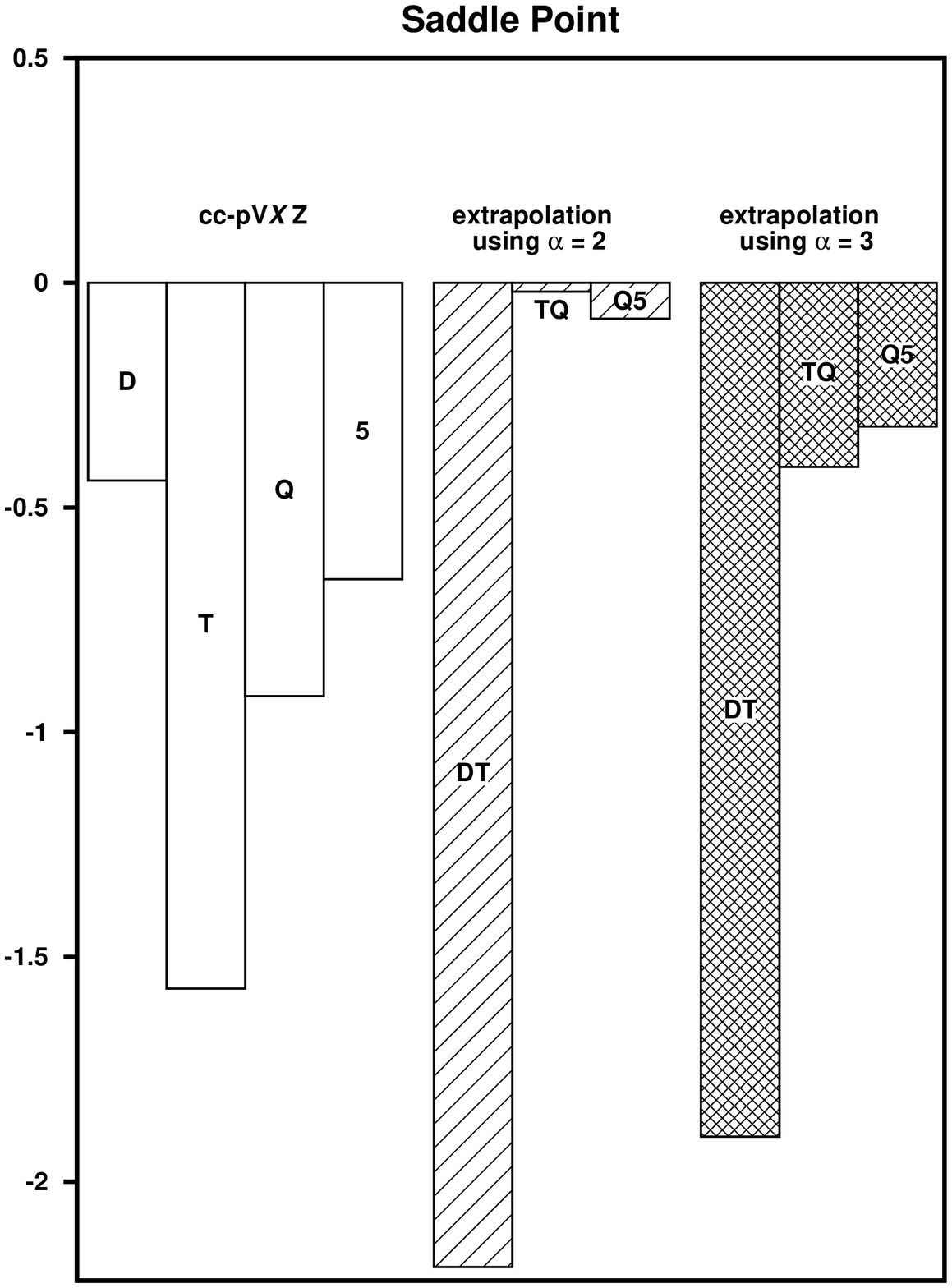}} & \hspace*{1.0cm}
{\includegraphics[scale=0.30,angle=0]{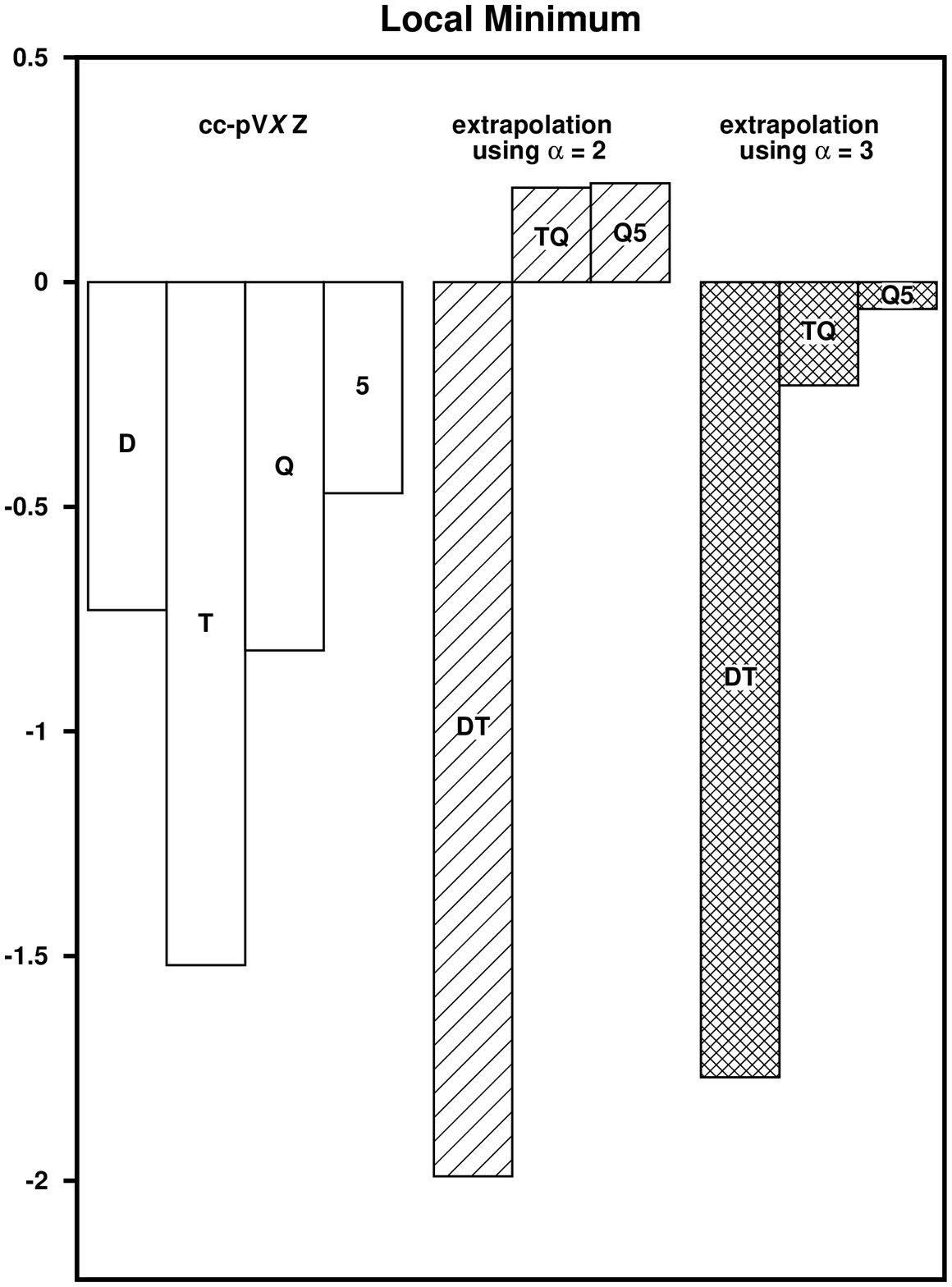}}\\%
& \hspace*{1.9cm}{\includegraphics[scale=0.39,angle=-90]{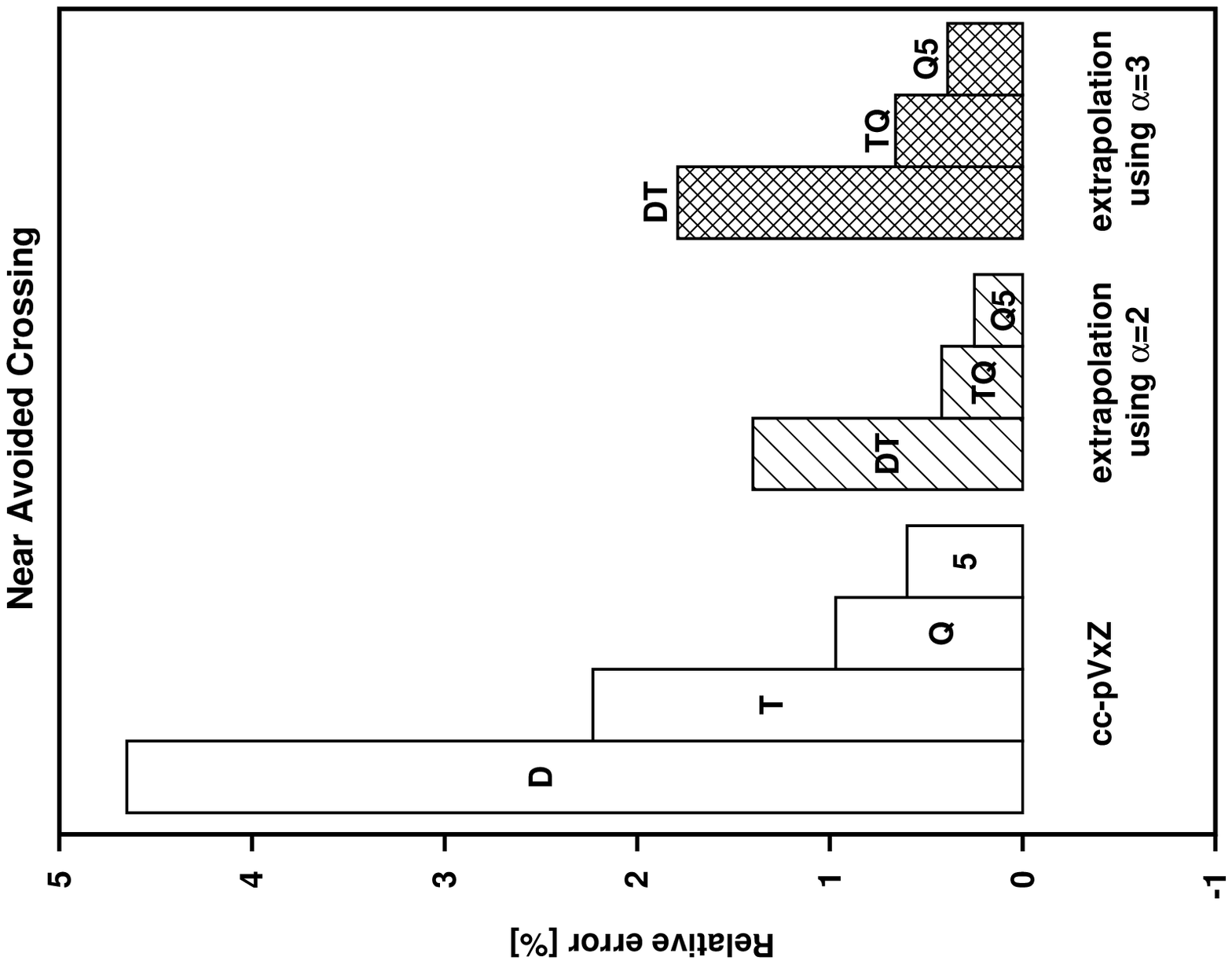}} & 
\end{tabular}
\end{center}
\caption{Relative percentage errors of the interaction energy at the
characteristic points (global minimum, saddle point, local minimum, and 
near the avoided crossing: $R$=5.5 bohr and $\theta=0^\circ$)
of the LiH--Li potential calculated at the CCSD(T) / cc-pV$X$Z-mid
level, where $X=$ D, T, Q, 5 and mid stands for the midbond-95 set.
The errors for the characteristic points obtained by extrapolating
the plain basis-set results with the two-point extrapolation formula
are also shown for $\alpha=2$ and $\alpha=3$. The errors were obtained
by comparison with the CCSD(T)-F12b / QZVPP results.}
\label{fig1}
\end{figure}

The relative percentage errors, $\Delta_{\rm F12b}$, are plotted in
Fig.~\ref{fig1} for both plain (non-extrapolated) and extrapolated
characteristic points. For the global minimum, the plain cc-pV$X$Z
results approach the basis-set limit from above and the convergence is
smooth and fast: the error is reduced by a factor of 2 to 3 for each
increment in $X$. The extrapolation accelerates the convergence: the
$(X-1)X$ extrapolated interaction energies have a quality at least that
of the plain cc-pV$(X+1)$Z results. Though the extrapolation with
$\alpha = 2$ seems to be more efficient than that with $\alpha = 3$ for
the DT and TQ cases, it actually overshoots the basis-set limit when
the Q and 5 cardinal numbers are used. More importantly, using
$\alpha=2$ leads to irregular behaviour: the Q5 extrapolation results
in a lower quality than the TQ extrapolation. In contrast,
extrapolation with $\alpha=3$, though slightly less efficient for low
cardinal numbers, exhibits highly systematic behaviour and leads to an
error as small as 0.01\% for the Q5 extrapolation.

Similar behaviour of the extrapolation schemes is observed for 
the point near the avoided crossing. Both extrapolations, with $\alpha=2$
and $\alpha=3$, converge smoothly towards the basis-set limit, 
but the convergence is not as fast as in the case of the global minimum.
In contrast to the global minimum, there is no problem here with
overshooting the basis-set limit. For each pair of cardinal numbers $(X-1)X$
the extrapolation with $\alpha=2$ gives results slightly more favourable
than using $\alpha=3$, with the smallest error of $0.19\%$ for the 
Q5 extrapolation.

For the saddle point and local minimum, the convergence of the
relative errors is not as smooth as for the global minimum: the
relative error for $X$=D is surprisingly small. This is obviously
accidental and does not reflect particularly high quality of the
cc-pVDZ basis set. Indeed, when the cc-pVDZ results are employed in
Eq.~(\ref{extrapol}), the extrapolation worsens the accuracy: the
errors for the DT extrapolation are much larger than the errors for
both the $X$=D and $X$=T plain results, independent of the value of the
$\alpha$ extrapolation parameter. Starting from $X$=T, the plain
results smoothly approach the basis-set limit, though the convergence
is clearly slower than in the case of the global minimum. The
extrapolation with $\alpha = 2$ is unsystematic and unpredictable, as
in the case of the global minimum, while that with $\alpha = 3$
smoothly approaches the basis-set limit. The errors of the Q5
extrapolation with $\alpha=3$ are $-0.49\%$ for the saddle point and
$-0.13\%$ for the local minimum.

Patkowski and Szalewicz \cite{Patkowski:2010} recently investigated Ar$_2$ with the CCSD(T)-F12
method. They found that the F12a and F12b variants \cite{ccf12} gave significantly different
results. They also concluded that, for Ar$_2$, calculations
with explicitly correlated functions
cannot yet compete with calculations employing extrapolation based on conventional orbital basis sets.
Indeed, while their orbital results converged smoothly towards the extrapolated results,
the CCSD(T)-F12a and CCSD(T)-F12b
results behaved erratically with respect to both the orbital and the extrapolated results. Table
\ref{tab0} shows that this is not the case for the
Li--LiH system. In our case the CCSD(T)-F12a and CCSD(T)-F12b results are quite similar and are
fully consistent with the plain and extrapolated results with conventional basis sets.
It should be stressed, however, that Ar$_2$ is bound mostly by dispersion forces,
while the main source of the bonding in Li--LiH is the induction energy, which is less sensitive to
the basis-set quality. This may at least partly explain the success of the CCSD(T)-F12 calculations
for Li--LiH.

Finally, it is important to note here that, while the interaction energy at the
characteristic points varies considerably with the basis set and
extrapolation method, the positions of the points (i.e., the distance
$R$ and angle $\theta$ at which the characteristic points occur) remain
practically unaffected by the choice of the basis set and extrapolation
scheme.

\begin{figure}
\vspace*{2cm}
\begin{center}
\begin{tabular}{c}
{\includegraphics[scale=0.65,angle=-90]{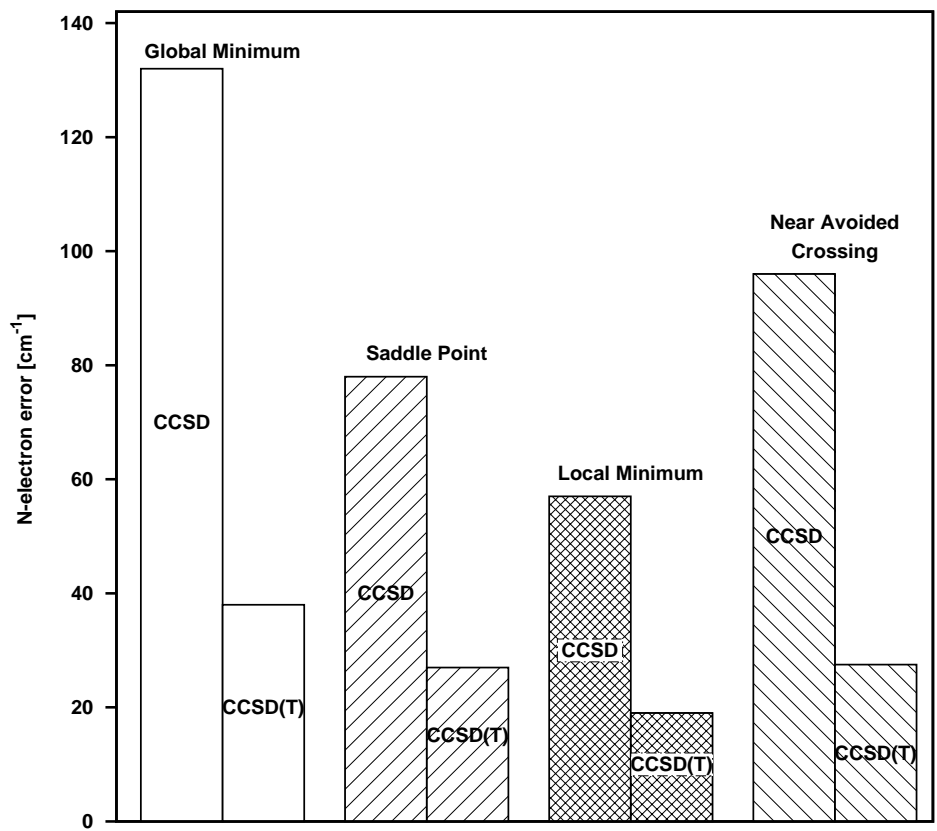}}
\end{tabular}
\end{center}
\caption{The $N$-electron error of the characteristic points of the
LiH--Li interaction potential calculated at the CCSD / cc-pVQZ and
CCSD(T) / cc-pVQZ levels of theory. The error was determined by
comparison with the FCI / cc-pVQZ interaction potential. The
$1\sigma_{\mathrm{LiH}}$ $1s_{\mathrm{Li}}$ frozen-core approximation
was used.
}
\label{fig2}
\end{figure}

To analyze the convergence of the CCSD and CCSD(T) models in the
$N$-electron space, Fig.~\ref{fig2} compares the characteristic points
(global minimum, saddle point, local minimum, and near the avoided crossing) 
of the Li--LiH
potential calculated at the CCSD / cc-pVQZ and CCSD(T) / cc-pVQZ levels
of theory with the characteristic points obtained at the FCI / cc-pVQZ
level. The $1\sigma_{\mathrm{LiH}}$ and $1s_{\mathrm{Li}}$ orbitals
were kept frozen in the calculations. As expected, the $N$-electron
error is reduced by a factor of 3 to 4 when the approximate triples
correction is included in the calculations. It can also be seen from
the figure that the global minimum is the most sensitive and the local
minimum is the least sensitive to the description of the electron
correlation.

\subsection{The best estimate of the ground-state Li--LiH potential energy surface}
\label{sub2} Because of the negligible one-electron error in the
CCSD(T)--F12 calculations and to the rather large basis set used in the
FCI / cc-pVQZ calculations, and assuming that the one-electron and
$N$-electron errors are approximately independent, the best estimate of
the ground-state interaction energy surface for the LiH-Li is
\begin{equation}
\label{best}
V^{\rm best} = V^{\rm CCSD(T)-F12} + \delta V_{\rm v-v}^{\rm FCI} + \delta V^{\rm FCI},
\end{equation}
where $V^{\rm CCSD(T)-F12}$ is the CCSD(T) basis-set limit energy
(i.e., the CCSD(T)-F12 result) and the FCI correction, $\delta V_{\rm v-v}^{\rm
FCI}$, is obtained by subtracting the CCSD(T) / cc-pVQZ
energy from the FCI / cc-pVQZ energy, both calculated in the
frozen-core approximation. The quantity $\delta V^{\rm FCI}$ accounts
for the last remaining correction (in the non-relativistic limit),
namely the effects of core-core and core-valence correlation in the FCI
/ cc-pVQZ calculations,
\begin{equation}
\delta V^{\rm FCI} = \delta V^{\rm FCI}_{\rm all-all} - \delta V^{\rm FCI}_{\rm v-v},
\end{equation}
where the subscript ``all'' refers to all electrons correlated.

The quantity $\delta V^{\rm FCI}$ is a measure of the uncertainty in
our best estimate $V^{\rm best}$. To estimate this, we may safely
assume that $\delta V^{\rm FCI}$ is at most as large as the
corresponding $\delta V^{\rm (T)}$,
\begin{equation}
\label{assumptionT}
\delta V^{\rm FCI} \le \delta V^{\rm (T)} = \delta V^{\rm (T)}_{\rm all-all} -
\delta V^{\rm (T)}_{\rm v-v},
\end{equation}
where
\begin{equation}
\delta V^{\rm (T)}_{\rm all-all}  = V^{\rm CCSD(T)}_{\rm all-all} - V^{\rm CCSD}_{\rm all-all}
\end{equation}
\begin{equation}
\delta V^{\rm (T)}_{\rm v-v}  = V^{\rm CCSD(T)}_{\rm v-v} - V^{\rm CCSD}_{\rm v-v},
\end{equation}
with $V^{\rm CCSD(T)}_{\rm all-all}$, $V^{\rm CCSD}_{\rm all-all}$, $V^{\rm
CCSD(T)}_{\rm v-v}$, and $V^{\rm CCSD}_{\rm v-v}$ denoting
interaction energies calculated at the CCSD(T) / cc-pVQZ  or CCSD /
cc-pVQZ level, correlating all electrons or using the frozen-core
approximation, as appropriate. As can be seen from Fig.~\ref{fig2}, the
differences between CCSD(T) and CCSD are, for the characteristic points
of the potential, 2 to 3 times larger (and for the rest of the
potential at least 1.5 times larger) than the differences between FCI
and CCSD(T). Eq.~(\ref{assumptionT}) is therefore actually a
conservative estimate for $\delta V^{\rm FCI}$. The root mean square
error for $\delta V^{\rm (T)} / \delta V^{\rm (T)}_{\rm all-all}$, over the whole
potential is $4.1\%$. We thus consider that our best estimate of the
ground-state interaction energy for LiH--Li, Eq.~(\ref{best}), has a
(conservative) total uncertainty of $5\%$ of the FCI correction
($\delta V^{\rm FCI}_{\rm v-v}$). The analysis of the Li--LiH potential in
the remainder of this paper is based on the interaction energies
obtained using Eq.~(\ref{best}), unless otherwise stated.

To justify our error estimation we have performed calculations with
all electron correlated at the FCI level for the set of characteristic points 
of the potential. Due to the immense memory requirements of the FCI
calculations with seven electrons we were able to apply the cc-pVDZ
basis set only. The FCI/cc-pVDZ results together with the  CCSD(T)/cc-pVDZ,
both with and without the frozen-core approximation, are presented in Table \ref{tabF}. 
The error in the FCI correction calculated with frozen core is
as small as 0.76 \% for the examined points. 
We may see that the approximation with the FCI
valence correction added to CCSD(T), Eq. (\ref{Pig}), reproduces the exact FCI results
with accuracy better than 1\% of the FCI correction
($\delta V^{\rm FCI}_{\rm v-v}$). 
This confirms  our estimate of 
$5\%$  uncertainty  in the FCI correction $\delta V^{\rm FCI}_{\rm v-v}$.  

\subsection{Features of the ground-state potential energy surface}
\label{sub3} In Table \ref{tab1} we have listed the characteristic
points of the potential energy surfaces of the ground state, which
correlates at long range with Li($^2$S) + LiH (${\rm X}\, ^1\Sigma^+$),
and the first excited state, which correlates at the long range with
Li($^2$P) + LiH (${\rm X}\, ^1\Sigma^+$). Both these states are of
$^2A^\prime$ symmetry in the $C_{s}$ point group. The latter is
included in Table \ref{tab1} since, as will be discussed in the next
section, it shows an avoided crossing with the ground-state potential
for the linear LiH--Li geometry. Table \ref{tab1}  shows that the
interaction potential for the ground state of Li--LiH is deeply bound,
with a binding energy of 8743 cm$^{-1}$ at the global minimum. The
global minimum is located at a skew geometry with $R_e$=4.40 bohr and
$\theta_e$=46.5$^\circ$, and is separated by a barrier around $R$=6.3
bohr and $\theta$=136.0$^\circ$ from a shallow local minimum at the
linear Li--LiH geometry. The local minimum is at $R$=6.56 bohr, with a
well depth of only 1623 cm$^{-1}$. The excited-state potential shows
only one minimum, at $R$=5.66 bohr, with a binding energy of 4743
cm$^{-1}$.

\begin{figure}
\vspace*{2cm}

{\includegraphics[scale=1.10,angle=0]{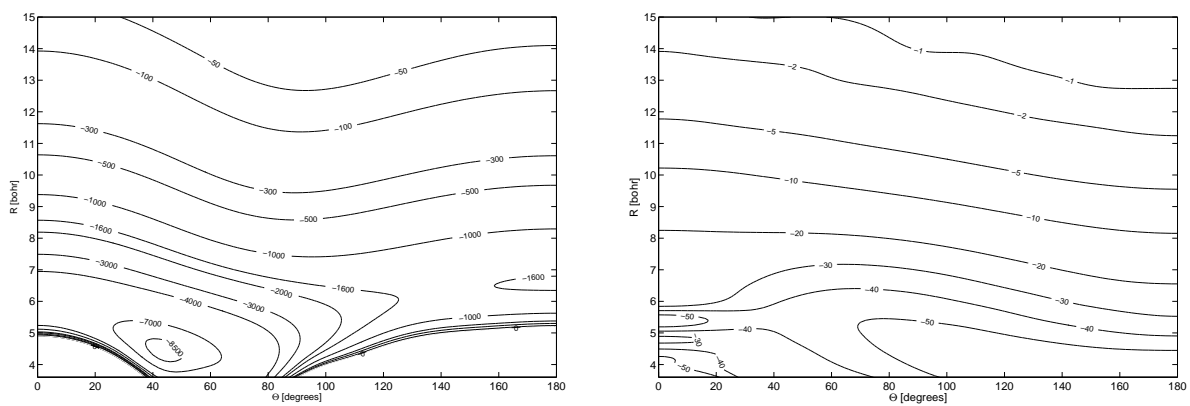}}\\
%\begin{tabular}{cc}
%%%\includegraphics[scale=0.27,angle=0]{Ground_full.eps}&\hspace*{-1.0cm}
%%%\includegraphics[scale=0.27,angle=0]{fci_corr.eps}
%\end{tabular}
\caption{Contour plots of the best {\em ab initio} potential for the
ground state of Li--LiH (left-hand panel), and of the full-CI
correction to the CCSD(T)--F12 potential (right-hand panel). Energies
are in cm$^{-1}$.} \label{fig3}
\end{figure}

A contour plot of the ground-state potential is shown in the left-hand
panel of Fig.\ \ref{fig3}, while the full-CI correction to the CCSD(T)
potential, $\delta V_{\rm v-v}^{\rm
FCI}$, is shown in the right-hand panel. The correction is very
small compared to the best potential. It amounts to 0.4\% around the
global minimum, and approximately 1\% at the local minimum. Thus, our
estimated error of the calculation, 5\% of the full-CI correction,
translates into 0.05\% error in the potential itself. We would like to
reiterate here that such a small error was achieved not only because
the interelectron distance was included explicitly in the {\em ab
initio} CCSD(T)--F12 calculations, but also because of the very small
valence-valence correlation beyond the CCSD(T) level. The smallness of
the valence-valence correlation beyond the CCSD(T) level is not so
surprising, since Li--LiH has only three valence electrons, and the
exact model for a three-electron system would be CCSDT, coupled-cluster
with single, double, and exact triple excitations \cite{ccsdt}. Our
results show that the triples contribution to the correlation energy
beyond the CCSD(T) model for the valence electrons is very small.

\begin{figure}
\vspace*{2cm}
\includegraphics[scale=0.70,angle=-90]{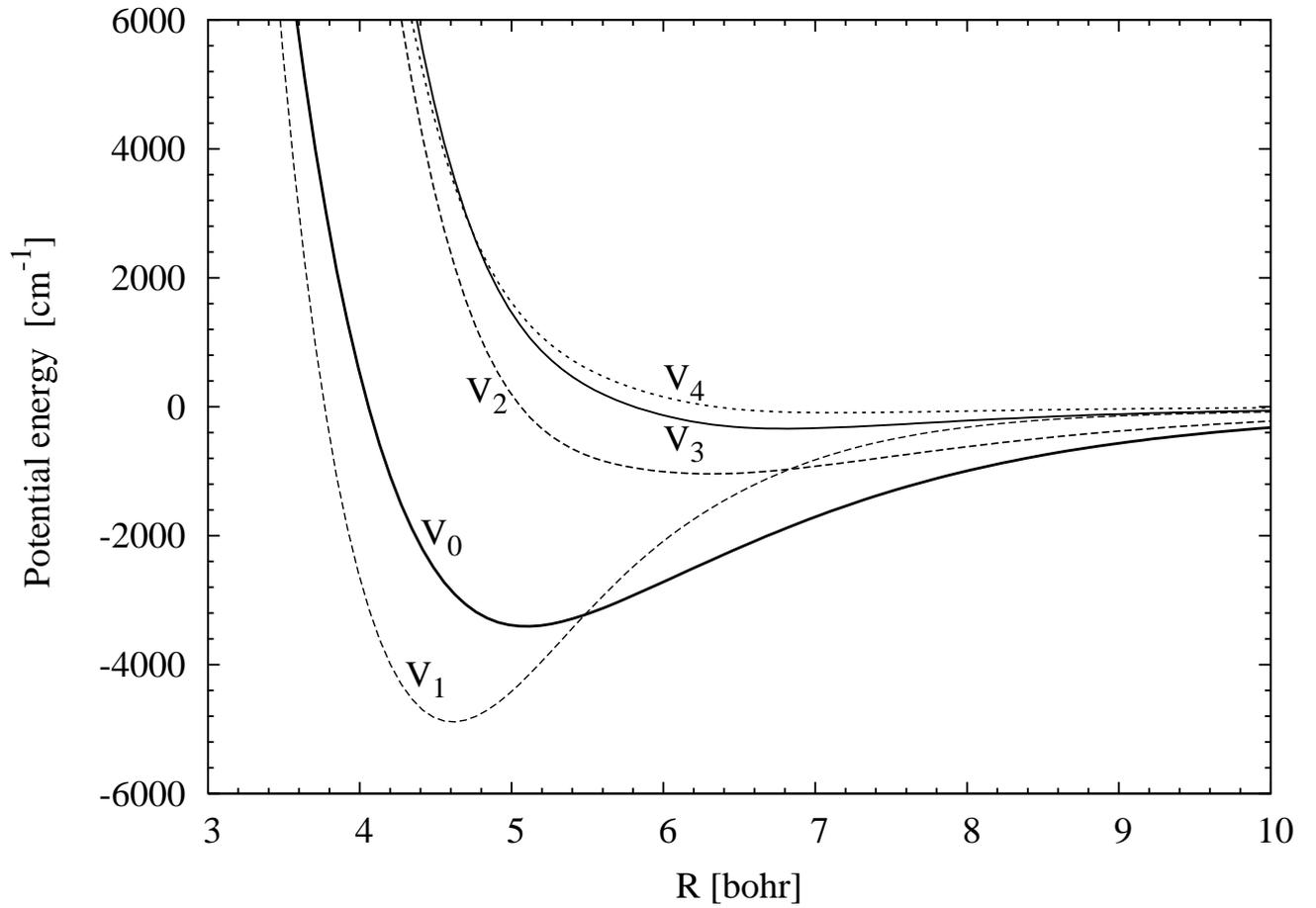}
\caption{The Legendre components $V_l(R)$ ($l=$ 0, 1, 2, 3, 4) of the
ground-state Li($^2$S) + LiH (${\rm X}\, ^1\Sigma^+$) interaction
potential, see Eq.~(\ref{anisotropy}).} \label{fig4}
\end{figure}

The potential for the ground state of Li--LiH is very strongly
anisotropic. This is easily seen in the left-hand panel of Fig.\
\ref{fig3}, and in Fig.\ \ref{fig4}, which shows the expansion
coefficients of the potential in terms of Legendre polynomials
$P_l(\cos\theta)$,
\begin{equation}
\label{anisotropy}
V(R,\theta) = \sum_{l=0}^{\infty} V_l(R) \, P_l(\cos\theta).
\end{equation}
Here, $V_0(R)$ is the isotropic part of the potential and
$\left\{V_l(R)\right\}_{l=1}^{\infty}$ is the set of anisotropic
coefficients. Fig.\ \ref{fig4} shows that, around the radial position
of the global minimum, $R$=4.36 bohr, the first anisotropic
contribution to the potential, $V_1(R)$, is far larger than the
isotropic term, $V_0(R)$. The higher anisotropic components, with $l=2,
3$, etc., contribute much less to the potential.

As mentioned above, calculations of collision dynamics at ultralow
temperatures require accurate values of the long-range potential
coefficients, Eq.\ (\ref{lr1}). Some important scattering properties,
such as the mean scattering length and the heights of centrifugal
barriers, are determined purely by the Van der Waals coefficients. The
calculated coefficients for Li--LiH are presented in Table \ref{tab2}.
Because of the large dipole moment of lithium hydride and the
relatively high polarizability of the lithium atom, the lowest-order,
most important, coefficients are dominated by the induction
contribution. For example, the induction part of $C_6^0$ and $C_6^2$ is
887 a.u., which accounts for 71\% of $C_6^0$ and 98\% of $C_6^2$.

\section{Interaction between the ground and excited states}
\label{sec3}
\subsection{Low-lying excited state potential, nonadiabatic coupling
matrix elements, and diabatic potentials} \label{sub4}

% The ground-state
% CCSD(T) calculations with {\sc molpro} were performed in a loop,
% starting from the smallest distance $R$ and utilizing the converged HF
% determinant and the CCSD amplitudes from the previous distance as
% starting points for the next $R$. However, with this strategy, we
% experienced some convergence problems at the linear LiH--Li geometry
% around $R$=5.6 bohr. We therefore repeated the calculations starting
% from large distances, decreasing towards $R$=5.6 bohr. Surprisingly,
% around 5.6 bohr the two sets of calculations converged to different
% solutions, as illustrated in the left-hand panel of Fig.\ \ref{fig5}.
% Curve $(a)$ was obtained in the calculations starting from small $R$,
% while curve $(b)$ was obtained in those starting from large $R$. It
% turned out that around 5.6 bohr the CCSD(T) calculations sometimes
% converged to an excited-state solution. This may happen since the
% coupled-cluster method is nonvariational, and in the case of
% quasi-degeneracy it can converge to excited-state solutions. Such
% behaviour of CCSD and other models has previously been observed for
% model systems such as H$_4$ \cite{jank1,jank2,jez}. Multiple solutions
% of the coupled-cluster equations have also been discussed by Kowalski
% and Jankowski \cite{kow1,kow2}.

\begin{figure}
\vspace*{2cm}
\begin{tabular}{cc}
%{\includegraphics[scale=0.30,angle=-90]{short_vs_long.eps}}  &
{\includegraphics[scale=0.50,angle=-90]{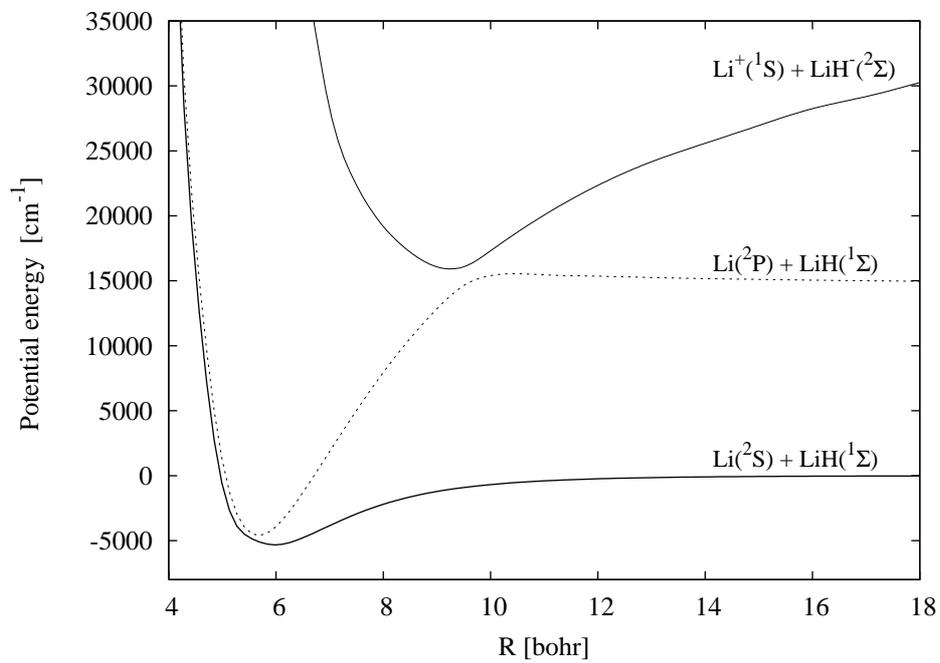}} \\%[5cm]
\end{tabular}
\caption{Ground-state and excited-state potentials at the linear geometry LiH--Li:
%The left-hand panel presents the convergence of the CCSD(T)
%calculations towards the excited-state solution. The right-hand panel
demonstration of  the ion-pair nature of the excited-state potential.}
\label{fig5}
\end{figure}

We encountered convergence problems with CCSD(T) calculations
at the linear LiH--Li geometry around $R$=5.6 bohr, due to the
presence of a low-lying excited state.
% The existence of a low-lying excited potential energy surface of
% $^2A^\prime$ symmetry was confirmed by performing calculations with the
% equation-of-motion coupled-cluster method with single and double
% excitations (EOM-CCSD) \cite{Monkhorst:77,Sekino:84,Stanton:93} in the orbital cc-pVQZ basis set, using
% the {\sc qchem} code \cite{qchem}.
%The results are shown in Fig.\ \ref{fig5} as dashed lines.
The excited state correlates with the
Li($^2$P)+LiH(X$^1\Sigma^+$) dissociation limit, but closer
investigation revealed that, at linear Li--HLi geometries near
the crossing with the ground state, it has ion-pair character,
Li$^+$($^1$S) + LiH$^-$($^2\Sigma$). The ion-pair state itself
has a crossing near $R=$ 9 bohr with the lowest $^2A^\prime$
state correlating with Li($^2$P) + LiH(X$^1\Sigma$). This is
shown schematically in
% the right-hand panel of
Fig.\ \ref{fig5}. Away from
linear Li--HLi geometries, the excited state has covalent character and
remains below the ion-pair state all the way to dissociation. The
avoided crossing between the ground state and the first excited state
is at $R$=5.66 bohr, which is near the minimum of the ground-state
potential at the linear geometry, and the energetic distance between
the two states at the avoided crossing is only 94 cm$^{-1}$.
% This suggests that the excited state might potentially have an impact on the
% Li+LiH scattering dynamics.

\begin{figure}
\begin{tabular}{c}
{\includegraphics[scale=0.70,angle=-90]{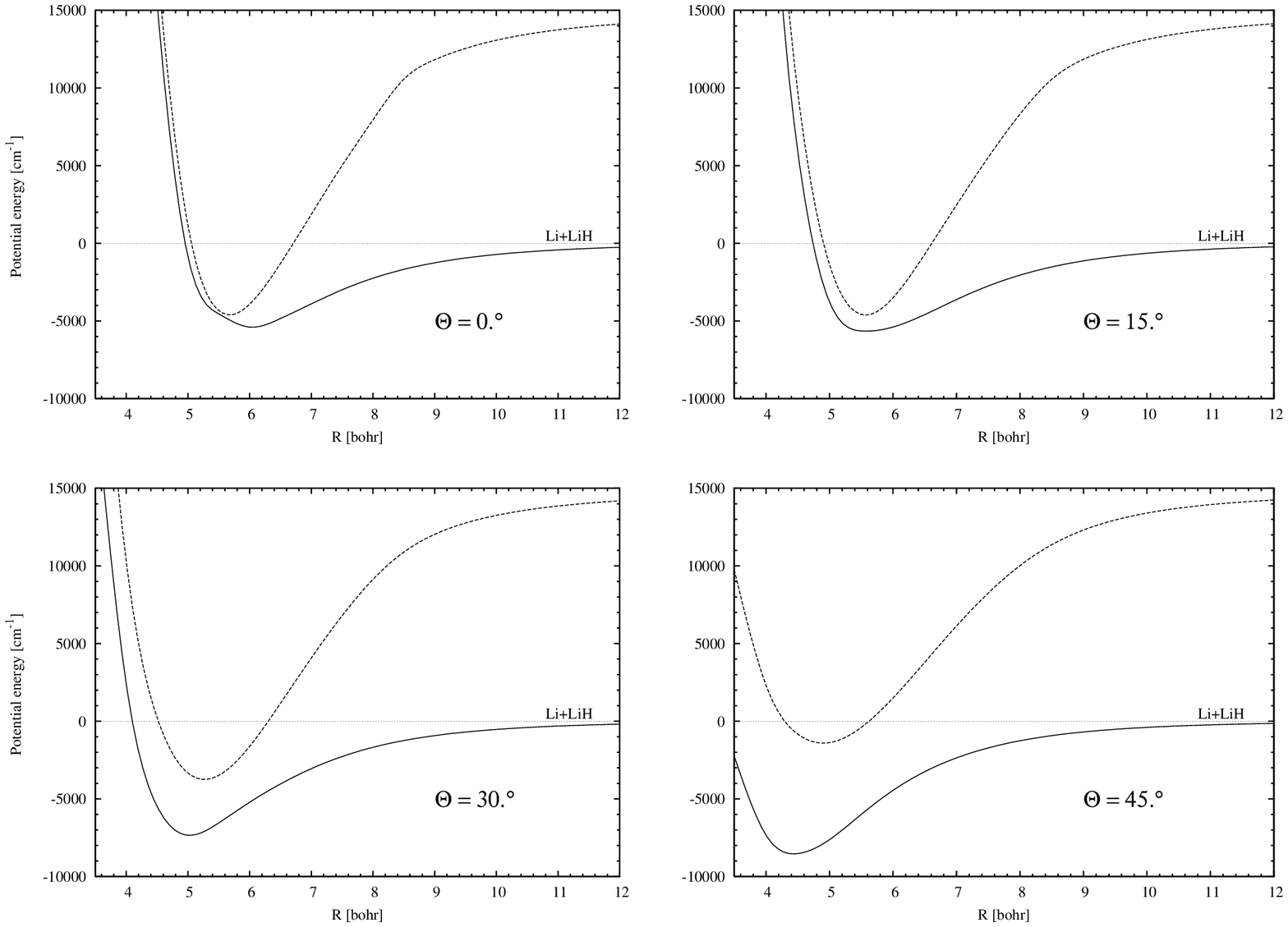}} \\%[10cm]
{\includegraphics[scale=0.70,angle=-90]{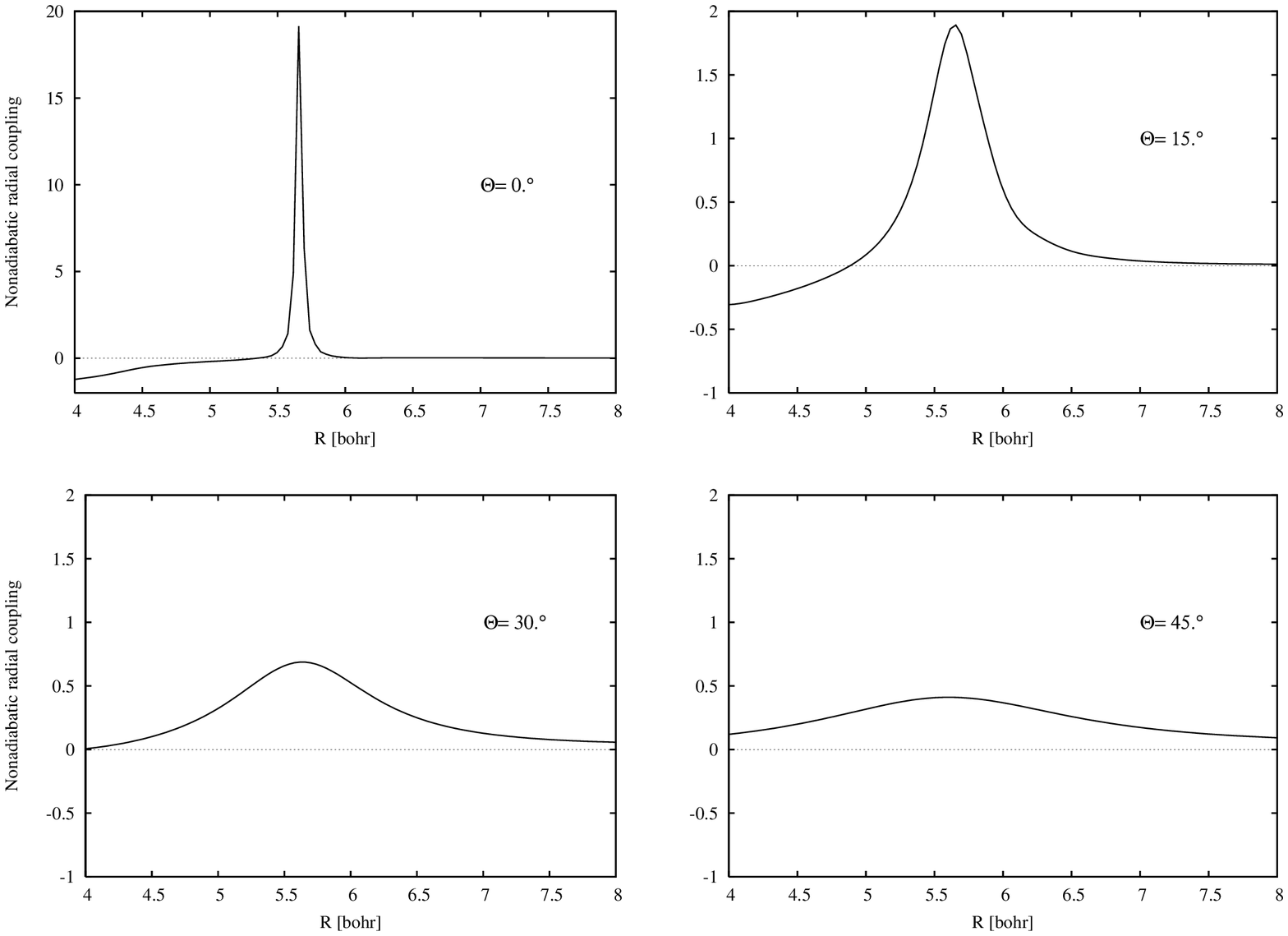}} \\%[5cm]
\end{tabular}
\caption{Cuts through the potential energy surfaces for the ground and
the first excited states of $^2A^\prime$ symmetry for selected values
of the angle $\theta$, and the corresponding radial nonadiabatic
coupling matrix elements as functions of the distance $R$.} \label{fig6}
\end{figure}

In order to investigate how far the excited state may affect the scattering dynamics,
we computed the full potential energy
surface for the excited state in question by means of
equation-of-motion coupled-cluster method with single and double
excitations (EOM-CCSD) \cite{Monkhorst:77,Sekino:84,Stanton:93}
implemented in the {\sc qchem} code \cite{qchem}, using the orbital cc-pVQZ basis set.
Cuts through the
ground-state and excited-state potential energy surfaces at selected
values of the angle $\theta$ are shown in Fig.\ \ref{fig6}. It may be
seen that it is only near the linear LiH--Li geometry that the two
states come very close together. If we distort the system from the
linear geometry, the excited state goes up in energy very rapidly, and
around the global minimum energy, $\theta\approx 45^\circ$, it is
almost 6000 cm$^{-1}$ above the ground state. The importance of the
possible interaction between the ground and excited states can be
measured by analyzing the (vectorial) nonadiabatic coupling  matrix
elements $\boldsymbol\tau_{12}$, defined as $\boldsymbol\tau_{12}=  \langle\Psi_1|\nabla \Psi_2\rangle$,
% \cite{baer,baer1}
% \begin{equation} \boldsymbol\tau_{12}=  \langle\Psi_1|\nabla \Psi_2\rangle,
% \label{nonad}
% \end{equation}
where $\nabla$ is the gradient operator of the position vector
$\bbox{R}$ and $\Psi_1$ and $\Psi_2$ are the wave functions of the two
lowest states. On the two-dimensional surface, we may define radial
$\tau_{12,R}=\langle\Psi_1|\partial \Psi_2/\partial R\rangle$ and
angular $\tau_{12,\theta}=\langle\Psi_1|\partial \Psi_2/\partial
\theta\rangle$ components of the vector $\boldsymbol\tau_{12}$.
We evaluated $\boldsymbol\tau_{12}$ for all $(R,\theta)$ geometries by
means of the multireference configuration interaction method limited to
single and double excitations (MRCI) \cite{WK88,Knowles:88}, using the
{\sc molpro} code \cite{MOLPRO2008}.
% To understand how the above
% expression measures the importance of the interaction between the two
% states in question, it is useful to rewrite it as
% \begin{equation} \boldsymbol\tau_{12}=  \frac{\langle\Psi_1|\nabla V|\Psi_2\rangle}{V_1-V_2},
% \label{nonad1}
% \end{equation}
% where $\nabla V$ is the gradient of the potential energy operator with
% respect to nuclear coordinates, and $V_1$ and $V_2$ are the potentials
% corresponding to the ground and excited states, respectively. Equation
% (\ref{nonad1}) clearly shows that
The nonadiabatic coupling is largest
%when the denominator is small, i.e.,
when the two states are very close in energy, as it can be seen in Fig.\ \ref{fig6}.
% Inspection of the four lower panels of Fig.\ \ref{fig6} confirm this.
While at $\theta=0^\circ$ the radial component of the
nonadiabatic coupling approaches the Dirac delta form near the crossing
point $R_{\rm ac}$,
% \begin{equation} \tau_{12,R}(R,\theta=0)\approx \frac{\pi}{2}\delta(R-R_{\rm ac}),
% \label{dd}
% \end{equation}
with increasing angle it becomes a broad function of approximately
Lorentzian shape. [Note the different scales on the vertical axes of
the different panels.]

The transformation from the adiabatic representation to a diabatic
representation may be expressed in terms of a mixing angle $\gamma$,
\begin{equation}
H_{1}=V_{2}\sin^2\gamma+ V_{1}\cos^2\gamma,
\; \; \; \;
H_{2}=V_{1}\sin^2\gamma+ V_{2}\cos^2\gamma,
\; \; \; \;
H_{12}=(V_{2}-V_{1})\sin\gamma \cos\gamma,
\label{adiab}
\end{equation}
where $V_{1}$ and $V_2$ are the ground-state and excited-state
adiabatic potentials, $H_1$ and $H_2$ are the diabatic potentials, and
$H_{12}$ is the diabatic coupling potential.
% In the diabatic picture,
% derivatives with respect to nuclear coordinates are eliminated from the
% equations that describe the nuclear motions, and are replaced by the
% coupling potential.
In principle, the mixing angle $\gamma$ may be
obtained by performing line integration of the nonadiabatic coupling
$\boldsymbol\tau_{12}$,
\begin{equation}
\gamma(\boldsymbol R)= \gamma(\boldsymbol R_0)+\int_{\boldsymbol R_0}^{\boldsymbol R}
\boldsymbol\tau_{12} \cdot  d{\boldsymbol l},
\label{gamma}
\end{equation}
where $\boldsymbol R_0$ is the starting point of the integration. For
polyatomic molecules, however, the mixing angle $\gamma$ obtained by
integrating this equation is non-unique due to the contributions from higher states.
% The source of the problem is
% the contributions from higher states, which make the integral
% path-dependent.
To circumvent the problem of path dependence, one may
 assume that we deal with an ideal two-state model.
% It can be shown \cite{baer} that in this case Eq.\ (\ref{gamma}) has a unique result,
% which can be written in the form
% \begin{equation}
% \gamma(R,\theta)= \gamma(R_0,\theta_0)+\int_{R_0}^R  \tau_{R}(R,\theta_0) \, {\rm d}R
%  + \int_{\theta_0}^\theta  \tau_{\theta}(R,\theta)\,  {\rm d}\theta .
% \label{gammaint}
% \end{equation}
% Equations (\ref{adiab}) to (\ref{gammaint}) are valid only for a
% two-state model.

In our case, however, the ion-pair surface
Li$^+$($^1$S)+LiH$^-$($^2\Sigma$) shows another crossing at small
angles and large distances, $\theta \le 15^\circ$ and $R\approx 9$
bohr, with another excited-state potential that correlates with the
Li$(^2{\rm P})$+LiH($^1\Sigma$) dissociation limit. Thus a third state
$\Psi_3$ comes into play and a two-state model is not strictly valid.
The energy of the first excited state goes up very rapidly
with the angle $\theta$, and at the same time the contribution of the
ion-pair configuration to the wave function of the first excited state,
$\Psi_2$, diminishes rapidly. Fortunately, the nonadiabatic coupling
matrix elements between the two lower states $\boldsymbol\tau_{12}$ and
between the two higher states $\boldsymbol\tau_{23}$ are well isolated.
The maximum  of $\boldsymbol\tau_{12}$ is separated from the maximum of
$\boldsymbol\tau_{23}$ by more than 4 bohr; the locations of the
crossing points between the surfaces for $\theta=0$ are shown in
% the right-hand panel of
Fig.\ \ref{fig5}. Moreover, the coupling
$\boldsymbol\tau_{13}$ between the ground state and the third state is
negligible over the whole configurational space. Thus, following the
discussion of Baer {\it et al.} \cite{Baer:2002} on the application of
the two-state model, we conclude that the necessary conditions are
fulfilled for the Li--LiH system. Due to the spatial separation of the
nonadiabatic couplings $\boldsymbol\tau_{12}$ and
$\boldsymbol\tau_{23}$, using the diabatization procedure based on the
two-state model is justified.
% This conclusion is further supported by
% the fact that we obtain calculated values of the mixing angle $\gamma$
% that depend very little on the integration path.
It is worth noting
that in our particular case we could not use the so-called
quasi-diabatization procedure \cite{Simah}, since it is not possible to
assign a single-reference wave function. This is due to the fact that
the excited state shows admixture from the ion-pair state.

% We remark also that the coupling $\boldsymbol\tau_{12}$ for geometries
% where the ion-pair structure (Li$^+$LiH$^-$) dominates the wave
% function of the first excited state $\Psi_2$ shows a sharp peak (Dirac
% delta-function-like) and is strictly localized around the avoided
% crossing. Conversely, for geometries where the wave function $\Psi_2$
% is dominated by the covalent structure Li$(^2{\rm P})$+LiH($^1\Sigma$),
% the coupling $\boldsymbol\tau_{12}$ is much broader and shows an
% $R^{-3}$ decay at large distances, with a constant proportional to the
% product of the dipole moment of LiH and the transition dipole moment of
% the Li atom between the ground $^2$S and first excited $^2$P states.

\begin{figure}
\vspace*{-0cm}
{\includegraphics[scale=1.30,angle=0]{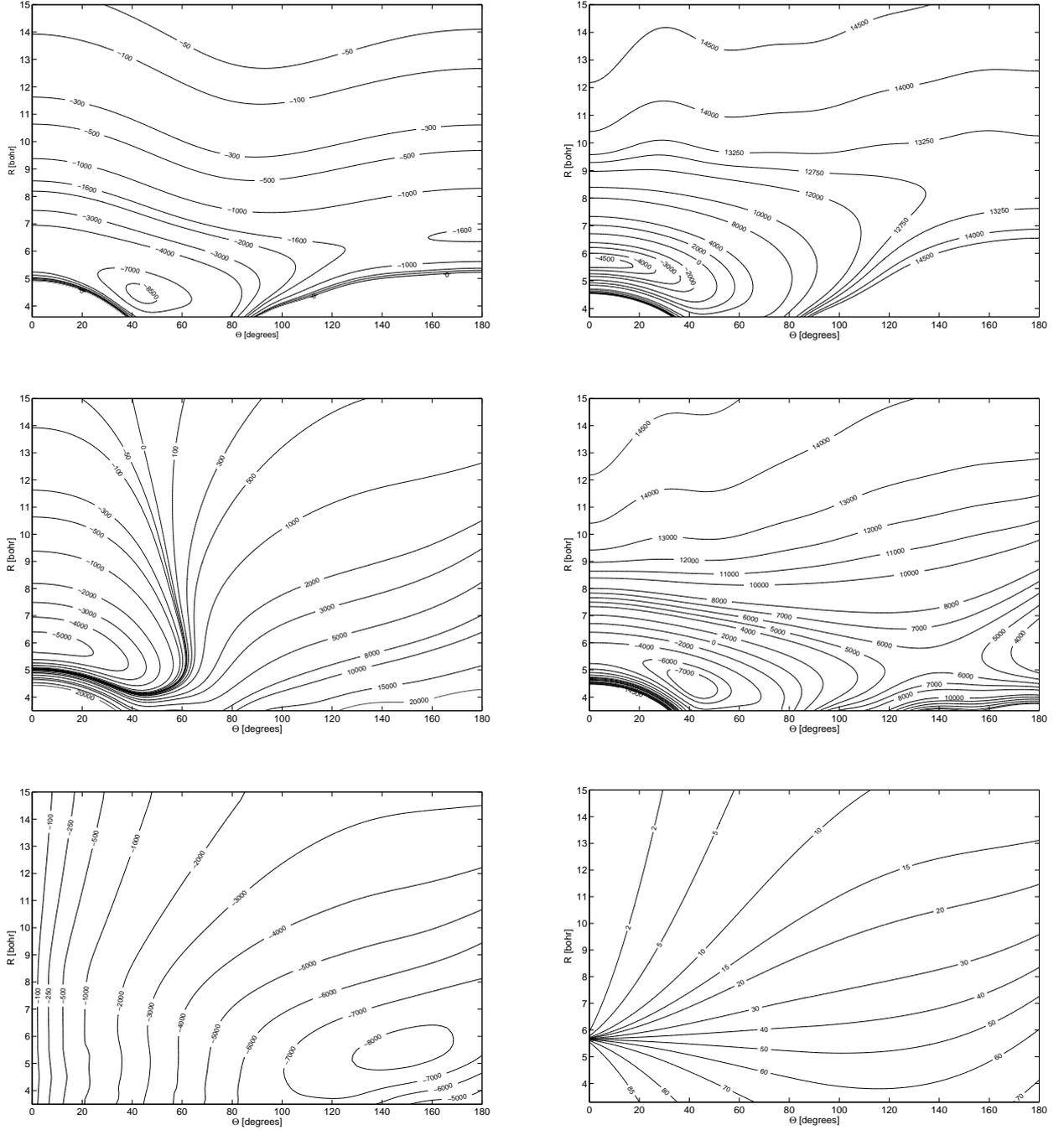}}\\
%\begin{tabular}{cc}
%\hspace*{-1cm}
%%{\includegraphics[scale=0.30,angle=0]{Ground_full.eps}}
%&\hspace*{-1cm}
%%{\includegraphics[scale=0.30,angle=0]{Exited_full.eps}} \\
%\hspace*{-1cm}
%%{\includegraphics[scale=0.30,angle=0]{Diabat_1.eps}}
%&\hspace*{-1cm}
%%{\includegraphics[scale=0.30,angle=0]{Diabat_2.eps}} \\
%\hspace*{-1cm}
%%\includegraphics[scale=0.30,angle=0]{Mixed.eps}
%&\hspace*{-1cm}
%%\includegraphics[scale=0.30,angle=0]{angle_diabat.eps}
%\end{tabular}
%\vspace*{0.0cm}
\caption{Adiabatic potentials (top two panels), diabatic potentials
(middle two panels), and coupling potential (bottom left-hand panel),
and the mixing angle $\gamma$ (bottom right-hand panel), as functions of
the geometrical parameters $R$ and $\theta$. Energies are in cm$^{-1}$.
} \label{fig7}
\end{figure}

As the starting point of the integration in Eq.\ (\ref{gamma}),
we chose $R=20$ bohr and $\theta=0^\circ$ and followed a radial
path along $\theta=0^\circ$ and subsequently angular paths at
constant $R$. The diabatic potentials were then generated
according to Eqs.\ (\ref{adiab}). Contour plots of the
adiabatic, diabatic, and coupling potentials, and of the mixing
angle $\gamma$, are presented as functions of $R$ and $\theta$
in Fig.\ \ref{fig7}. We consider first the mixing angle
$\gamma$, which is plotted in the bottom right-hand panel of
Fig.\ \ref{fig7}. As expected, the mixing angle shows an
accumulation point at $\theta=0^\circ$ at a distance $R$
corresponding to the closely avoided crossing between the
ground and excited states.
% A second
% accumulation point exists for the linear Li--LiH geometry, but it is
% completely unimportant for the dynamics, since it appears in the highly
% repulsive part of the potential at $R=1.66$ bohr.
For $\theta=180^\circ$, the mixing angle is non-negligible,
even at large distances. The coupling potential $H_{12}$
vanishes quite slowly with distance $R$, as $R^{-3}$. For the
coupling between the ground and ion-pair states, this
long-range decay is exponential, because of the different
dissociation limits of the two surfaces. As expected, at large
distances the two diabatic surfaces approach the respective
adiabatic surfaces. The diabatic surface that correlates
asymptotically with the excited-state Li($^2$P)+LiH surface has
an important contribution from the ground-state adiabatic
potential only inside the avoided crossing and at small angles
$\theta$. The diabatic surface that correlates asymptotically
with the ground state resembles the ground-state adiabatic
surface rather less closely, especially at large values of
$\theta$.
%The coupling potential
%tends to zero in the region of strong interaction between the ground
%and excited states, while it is largest (in absolute value) around
%$R=5.7$ bohr and $\theta=150^\circ$, i.e., in a region where the two
%diabatic potentials are highly repulsive. Physically, this means that
%the coupling potential is small in the region of the avoided crossing,
%and large in the region where the ground-state and excited-state
%potentials are well separated. We can thus conclude that nuclear
%dynamics will take place mostly on the lowest adiabatic surface.
The coupling between the diabatic states is small over a
significant region of $\theta$ and LiH bond length $r$ in the
vicinity of the crossing. Physically, this means that the
dynamics will be strongly nonadiabatic in this region, and to
take this rigorously into account would require a full
two-state treatment of the dynamics. However, there are no open
channels that involve the second surface, and any collisions
that cross onto it must eventually return to the original
surface. Its effect in collision calculations will therefore be
at most to cause a phase change in the outgoing wavefunction.

\subsection{Conical intersection}
\label{sub5} It is well known that potential energy surfaces for
homonuclear triatomic systems composed of hydrogen \cite{h3} or lithium
atoms \cite{li3} show conical intersections at equilateral triangular
geometries. Analogous behaviour may be expected for Li$_2$H, at
geometries where the two lithium atoms are equivalent, i.e., $C_{\rm
2v}$ geometries. Thus far, our discussion of the potential for Li--LiH
has been restricted to two dimensions with the bond length of the LiH
molecule fixed at its equilibrium value, and no conical intersection
was observed. However, if we start to vary the bond length of the LiH
molecule, conical intersections show up immediately.

\begin{figure}
\hspace*{-1cm}

{\includegraphics[scale=1.30,angle=0]{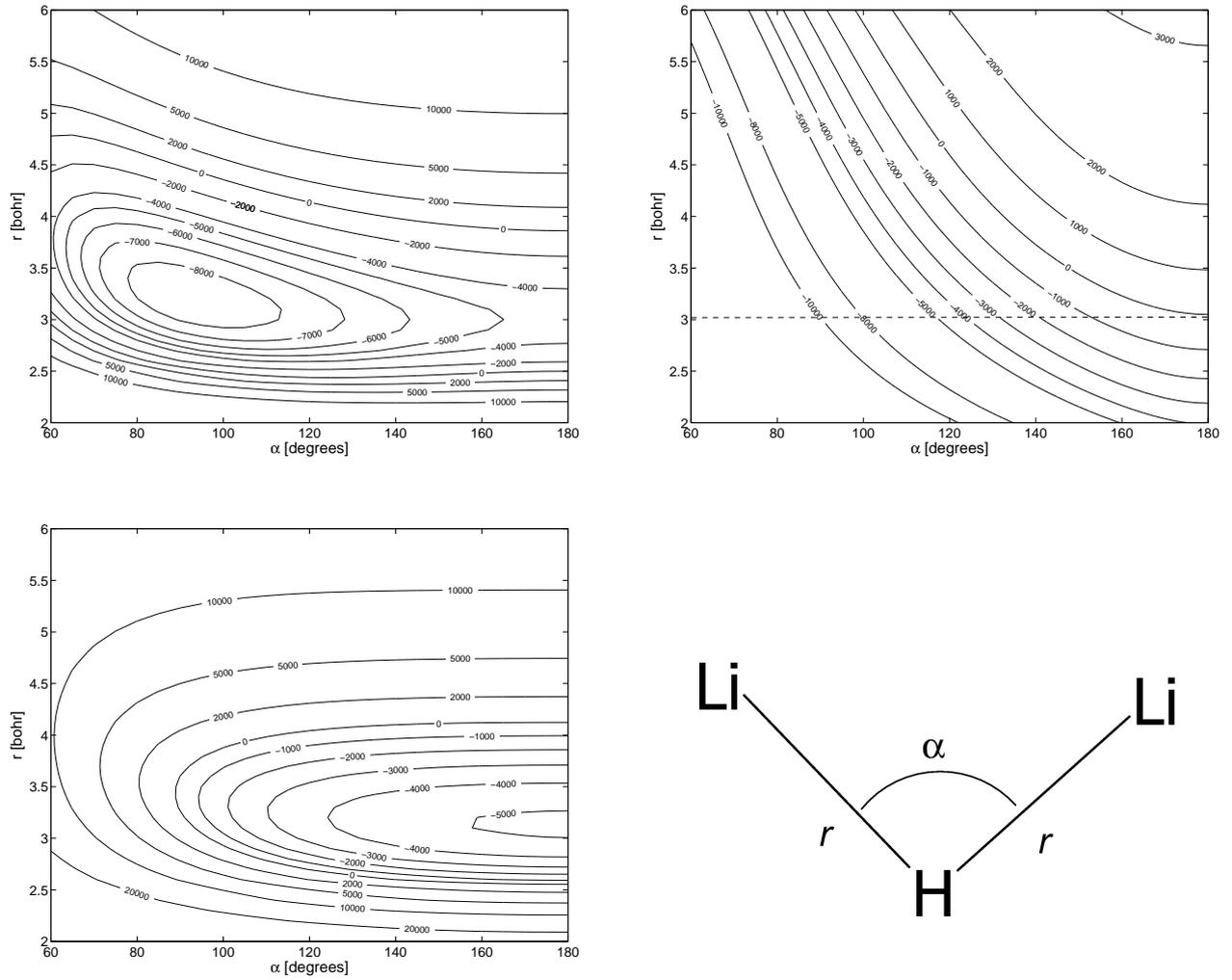}}\\
%\begin{tabular}{cc}
%\hspace*{-1cm}
%%{\includegraphics[scale=0.30,angle=0]{A1_plot.eps}}
%&\hspace*{0.3cm}
%%{\includegraphics[scale=0.30,angle=0]{Diff_cor.eps}} \\
%\hspace*{-1cm}
%%{\includegraphics[scale=0.30,angle=0]{B2_plot.eps}}
%&\hspace*{0.3cm}\vspace{1cm}
%%{\includegraphics[scale=0.25,angle=0]{eee1.eps}} \\
%\end{tabular}
\caption{Potential energy surfaces for the $^2$A$_1$ (top left-hand panel) and $^2$B$_2$ (bottom
left-hand panel)
states of Li$_2$H in $C_{\rm 2v}$ symmetry, and the difference between
them (top right-hand panel), which is zero along the seam of conical intersections. Also shown
is the coordinate system used for $C_{\rm 2v}$ geometries. Energies are
in cm$^{-1}$.} \label{fig8}
\end{figure}

At $C_{\rm 2v}$ geometries, with the two LiH bond lengths equal, there
are two low-lying electronic states, of $^2$A$_1$ and $^2$B$_2$
symmetries, that cross each other as a function of the internuclear
coordinates. Fig.\ \ref{fig8} shows contour plots of the two potential
energy surfaces and of the difference between them, and the top panel
of Fig.\ \ref{fig9} summarizes some key features of the surfaces.
The $^2$A$_1$ state has a minimum energy of $-8825$ cm$^{-1}$ at
$r$(LiH) = 3.22 bohr and an Li-H-Li angle of $95^\circ$. MRCI
calculations with all coordinates free to vary confirm that this is
indeed the absolute minimum geometry.
There is also a  saddle point
on the $^2$A$_1$ surface at a linear H-Li-H geometry with $r$(LiH) =
3.04 bohr and an energy of $-4992$ cm$^{-1}$, which is a minimum in $D_{\infty{\rm h}}$ symmetry.
The $^2$B$_2$ state has a
minimum energy of $-5136$ cm$^{-1}$ at $r({\rm LiH})=$ 3.17 bohr at a
linear Li-H-Li geometry. The $^2$A$_1$ saddle point and $^2$B$_2$ linear minimum have
symmetries $^2\Sigma_g^+$ and $^2\Sigma_u^+$ respectively in
$D_{\infty{\rm h}}$ symmetry, but mix and distort if the constraint on
the LiH bond lengths is relaxed, to form a $^2\Sigma^+$ state in
$C_{\infty{\rm v}}$ symmetry with a minimum at a linear geometry with
$r$(LiH) distances of 3.00 and 3.33 bohr and an energy of $-5323$
cm$^{-1}$. Even this is a saddle point with respect to bending on the
full potential surface in $C_s$ symmetry.

The $^2$A$_1$ and $^2$B$_2$ states are of different symmetries at
$C_{\rm 2v}$ geometries, but both are of $^2$A$^\prime$ symmetry when
the geometry is distorted from $C_{\rm 2v}$ to $C_{\rm s}$ symmetry.
The two states therefore mix and repel one another at geometries where
the two LiH bond lengths are different, but a seam of conical
intersections runs along the line where the energy difference is zero
at $C_{\rm 2v}$ geometries.

The fixed LiH distance used in previous sections ($r=3.014$ bohr, shown
as a dashed line on the figure) keeps the $^2$A$_1$ surface just below
the $^2$B$_2$ surface. However, if we allow for the vibrations of LiH,
the seam of conical intersections becomes accessible at near-linear
LiH--Li geometries, where the zero of the energy difference appears for
an Li--H distance only slightly larger than 3.014 bohr. At non-linear
geometries the seam quickly moves to Li--H distances far outside the
classical turning points of the ground vibrational level of free LiH,
which are 2.72 and 3.35 bohr.

%This means that the conical intersection
%is unlikely to play a significant role in the Li--LiH collision
%dynamics at low and ultralow energies.

It is interesting to compare the features of the conical
intersections in Li$_2$H with those in other triatomic
molecules formed from Li and H atoms: LiH$_2$, Li$_3$ and
H$_3$. In the case of LiH$_2$, the seam of intersections occurs
at highly bent $C_{2v}$ geometries with an angle between the
two Li-H bonds of approximately $30^\circ$ and arises from
degeneracy between the surfaces of A$_1$ and B$_2$ symmetry.
The global minimum of B$_2$ symmetry is located at $r({\rm
LiH})=3.23$ bohr and a bond angle H--Li--H of $28^\circ$
\cite{Yarkony:Li:1998}. This contrasts with Li$_2$H, where the
minimum of B$_2$ symmetry is at a linear Li--H--Li
configuration. The energy of the lowest point on the seam of
intersections is about 9000 cm$^{-1}$ above the
Li($^2$S)+H$_2(X^1\Sigma_g)$ threshold, so that it is
irrelevant for low and medium-energy collisions between H$_2$
and Li in their ground states, though it is important for
quenching of Li($^2$P) by H$_2$
\cite{Yarkony:Li:1998,Martinez:1997}.

The conical intersections for the doublet states of Li$_3$ and
H$_3$ occur at equilateral triangular geometries, where the
ground state is doubly degenerate and has symmetry
$^2$E$^\prime$ in the $D_{3h}$ point group. In the case of
H$_3$, the lowest-energy point on the seam is located at an
energy more than 20000 cm$^{-1}$ above the
H($^2$S)+H$_2(X^1\Sigma_g)$ threshold, so that nonadiabatic
effects are negligible in H+H$_2$ collisions \cite{Chu:2009},
although the conical intersection also produces geometric phase
effects \cite{Juanes-Marcos:2005}. For Li$_3$, the energetics
are essentially different. The lowest-energy point on the seam
is around 4000 cm$^{-1}$ below the
Li($^2$S)+Li$_2(X^1\Sigma_g)$ threshold and only 500 cm$^{-1}$
above the $C_{2v}$ global minimum \cite{Varandas:1998}. This is
likely to produce considerable nonadiabacity in collisions of
Li$_2$ with Li.

To conclude, in all the triatomic molecules formed from H and
Li there are seams of crossings that occur at configurations of
the highest possible symmetry, either $C_{2v}$ or $D_{3h}$. For
Li$_3$ and Li$_2$H the conical intersections are accessible
during atom-molecule collisions, while for H$_3$ and LiH$_2$
nonadiabatic processes are unimportant if the colliding
partners are in their ground states and have relatively low
kinetic energy.

\section{Reaction channels}
\label{sec5} 
\begin{figure}
\vspace*{2cm}
{\includegraphics[scale=0.30,angle=-90]{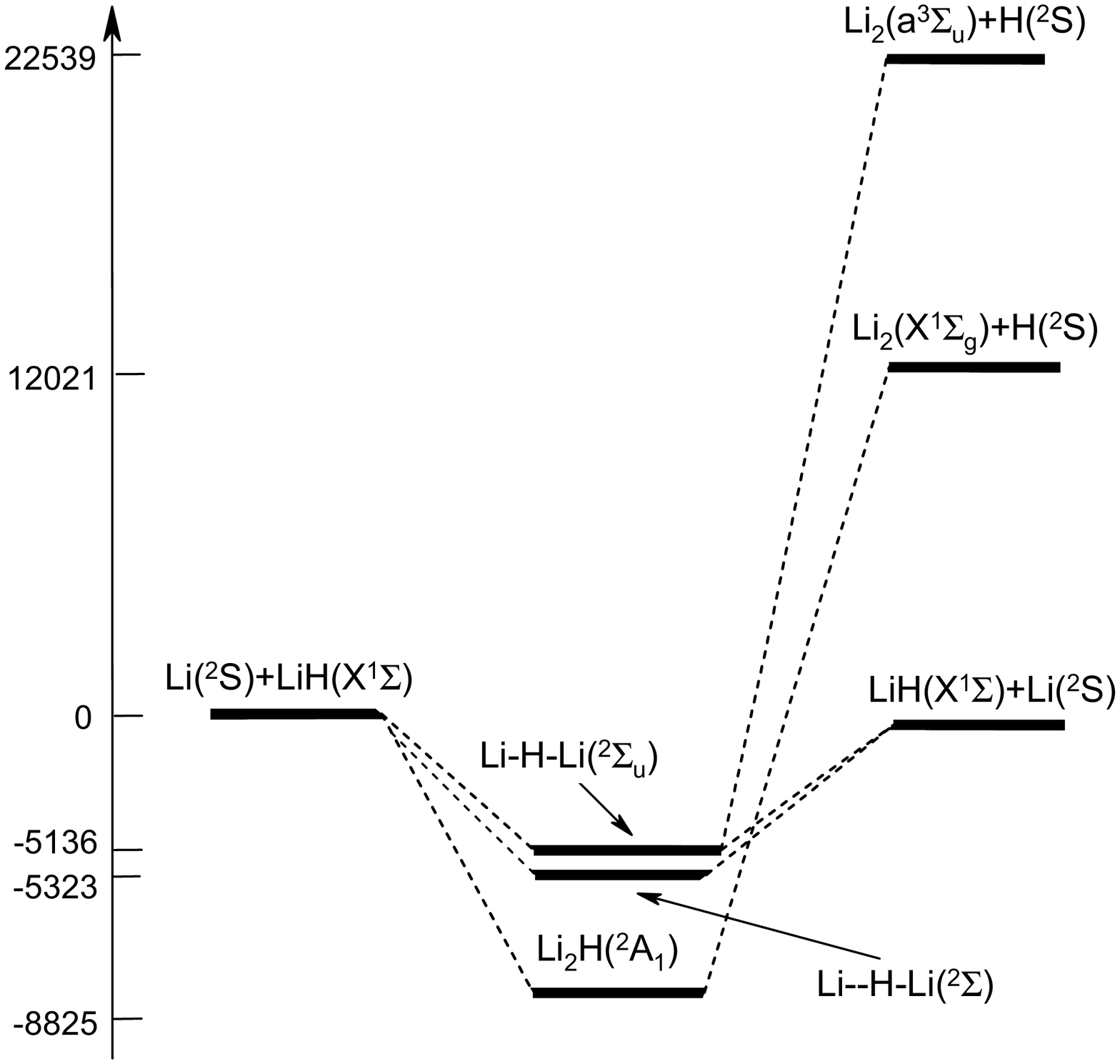}}\\
{\includegraphics[scale=1.10,angle=0]{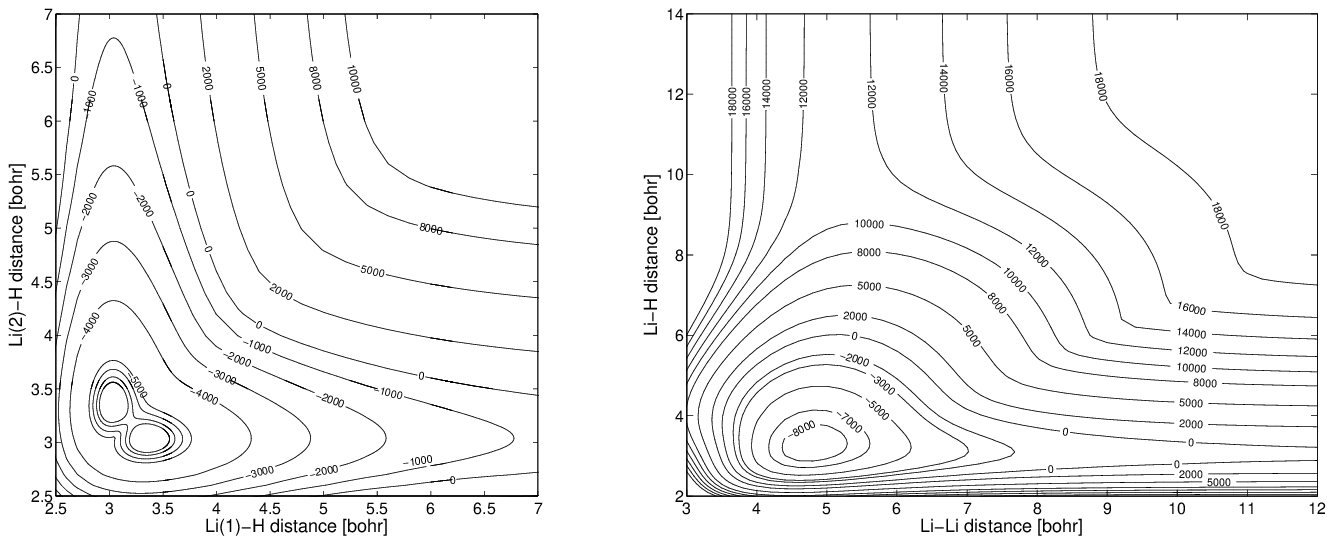}}
%\begin{tabular}{cc}
%%\resizebox{95mm}{!}{\includegraphics[scale=0.25,angle=0]{r2.eps}}&\hspace*{1.50cm}
%%\resizebox{80mm}{!}{\includegraphics[scale=0.25,angle=0]{r4.eps}}
%\end{tabular}
\caption{Schematic representation of the possible
reaction pathways in collisions of the lithium atom with the lithium
hydride molecule (upper panel), and two-dimensional cuts of the reactive
potential energy surfaces for the exchange (left-hand panel) and
insertion (right-hand panel) reactions. Energies are in cm$^{-1}$.}
\label{fig9}
\end{figure}

Several reaction channels exist that might affect
sympathetic cooling \cite{pz3} in Li+LiH. These are the exchange
reaction,
\begin{equation} {\rm LiH} + {\rm Li} \rightarrow {\rm Li} + {\rm HLi}
\label{exchange}
\end{equation}
and two insertion reactions,
\begin{equation} {\rm LiH} + {\rm Li} \rightarrow {\rm Li_2} + {\rm H},
\label{ins}
\end{equation}
producing Li$_2$(X$^1\Sigma_g^+$) and Li$_2$($a^3\Sigma_u^+$) plus a
ground-state H atom. The energetic location of the entrance and exit
channels of these reactions, as well as those of the potential minima
for linear and $C_{\rm 2v}$ geometries, are shown in the upper panel of
Fig.\ \ref{fig9}. The insertion reactions are highly endothermic, with
an energy difference between the entrance and exit channels of the
order of 12000 cm$^{-1}$ and 22500 cm$^{-1}$ for
Li$_2$(X$^1\Sigma_g^+$)+H and Li$_2$($a^3\Sigma_u^+$)+H, respectively.

To make the discussion more quantitative, Fig.\ \ref{fig9} also shows
two-dimensional plots of the energy as functions of the internal
coordinates. For the exchange reaction, we held Li--H--Li at linear
geometries and varied the distances from the two lithium atoms to the
hydrogen atom. For the insertion reaction, Li--Li--H was kept bent,
with the angle $\angle{\rm (HLi1Li2)}$ held constant at the $C_{\rm
2v}$ equilibrium value 42.5$^\circ$, while the Li--Li and Li--H
distances were varied. To make the plots consistent with the
correlation diagram shown on the upper panel, the zero of energy was
fixed at that of Li--LiH separated to infinite distance with the Li--H
bond length fixed at the monomer equilibrium value.

Let us consider the exchange reaction first. The two-dimensional cut
through the potential energy surface is presented in the left-hand
panel of Fig.\ \ref{fig9}. The potential energy surface of linear
Li$_2$H has two equivalent minima with an energy of $-5323$ cm$^{-1}$,
separated by a small barrier 187 cm$^{-1}$ high. The linear minima are
in any case substantially above the absolute minimum (8825 cm$^{-1}$),
so this small barrier will have no important effect on the collision
dynamics. The exchange reaction produces products that are
indistinguishable from the reactants, so reactive collisions cannot be
distinguished from inelastic collisions experimentally (unless the two
Li atoms are different isotopes).

An analogous two-dimensional cut through the potential energy surface
corresponding to Li$_2$(X$^1\Sigma_g$)+H products is presented in the
right-hand panel of Fig.\ \ref{fig9}. The plot illustrating the
reaction to form Li$_2$($a^3\Sigma_u$)+H products is not reported, as
the reaction is even more endothermic. The surface includes the
absolute minimum at an energy of $-8825$ cm$^{-1}$. The entrance
channel for this reaction corresponds to an Li--H distance of 3.014
bohr at large Li--Li distance, while in the exit channel the Li--Li
distance is approximately 5.05 bohr when the Li--H distance is very
large. However, this reaction cannot occur at low collision energies.

\section{Summary and Conclusions}
\label{sec6} In the present paper, state-of-the-art {\em ab initio}
techniques have been applied to compute the ground-state potential
energy surface for Li--LiH in the Born-Oppenheimer approximation. The
interaction potential was obtained using a combination of the
explicitly-correlated unrestricted coupled-cluster method with single, double, and
approximate noniterative triple excitations [UCCSD(T)--F12] for the
core-core and core-valence correlation, with full configuration
interaction for the valence-valence correlation. The main results of
this paper can be summarized as follows:
\begin{enumerate}
\item The Li--LiH system is strongly bound: if the LiH bondlength
    is held fixed at the monomer equilibrium distance of 3.014
    bohr, the potential energy surface has a global minimum 8743
    cm$^{-1}$ deep at a distance $R$=4.40 bohr from the lithium
    atom to the center of mass of LiH, and a Jacobi angle
    $\theta=46.5^\circ$. It also shows a weak local minimum 1623
    cm$^{-1}$ deep at the linear Li--LiH geometry for $R$=6.56
    bohr, separated from the global minimum by a barrier at
    $R$=6.28 bohr and $\theta=136^\circ$. If the LiH bond length is
    allowed to vary, the potential minimum is at a depth of 8825
    cm$^{-1}$, at a $C_{\rm 2v}$ geometry with LiH bond length of
    3.22 bohr and an Li-H-Li angle of $95^\circ$.

\item The full-CI correction for the valence-valence correlation to
    the explicitly correlated CCSD(T)--F12 potential is very small.
    The remaining error in our calculations is due to the neglect
    of the core-core and core-valence contributions, and is
    estimated to be of the order of 0.05\% of the total potential.

\item To evaluate the performance of the conventional
    orbital electron-correlated methods, CCSD and CCSD(T),
    calculations were carried out using
    correlation-consistent polarized valence $X$-tuple zeta
    basis sets, with $X$ ranging from D to 5, and a very
    large set of mid-bond functions. Simple two-point
    extrapolations based on the single-power laws $X^{-2}$
    and $X^{-3}$ for the basis-set truncation error
    reproduce the CCSD(T)--F12 results for the
    characteristic points of the potential with an error of
    0.49\% at worst.

\item The potential for the ground state of Li--LiH is strongly
    anisotropic. Around the distance of the global minimum, the
    isotropic potential $V_0(R)$ is almost two times smaller than
    the first anisotropic contribution $V_1(R)$. Higher anisotropic
    components, with $l=2, 3$, etc., do not contribute much to the
    potential.

%\item The origins of the bonding at the characteristic points of
%    the potential were analyzed using Symmetry-Adapted Perturbation
%    Theory (SAPT). This showed that the global minimum corresponds
%    to the maximum attraction and maximum repulsion  (in absolute
%    value), while the opposite is true for the local minimum. In
%    both cases the induction interaction between the strongly
%    dipolar LiH molecule and the highly polarizable Li atom
%    provides the main attractive component, compensated by large
%    repulsion contributions due to the first-order exchange
%    interaction and the second-order exchange-induction
%    interaction.

\item At the linear LiH--Li geometry, the ground-state
    potential shows a close avoided crossing with the first
    excited-state potential, which has ion-pair character
    around the avoided crossing point. The full potential
    energy surface for the excited state was obtained with
    the equation-of-motion method within the framework of
    coupled-cluster theory with single and double
    excitations. The excited-state potential has a single
    minimum 4743 cm$^{-1}$ deep for the linear LiH--Li
    geometry at $R$=5.66 bohr. The energy difference
    between the ground and excited states at the avoided
    crossing is only 94 cm$^{-1}$. An analysis of the
    nonadiabatic coupling matrix elements suggests that
    dynamics in the vicinity of the avoided crossing will
    have nonadiabatic character.

%    the interaction between the ground and excited
%    states should have little impact on the collision dynamics,
%    because the couplings are important only near the linear
%    geometry. A slight distortion from the linear configuration
%    makes the couplings negligibly small. Further analysis of the
%    diabatic potentials confirms this analysis.

\item When stretching the LiH bond in the Li--LiH system, a seam of
    conical intersections appears for $C_{\rm 2v}$ geometries,
    between the ground state of $^2$A$_1$ symmetry and an excited
    state of $^2$B$_2$ symmetry. At the linear LiH--Li geometry,
    the conical intersection occurs for an Li--H distance which is
    only slightly larger than the equilibrium distance of the LiH
    monomer, but for significantly non-linear geometries it moves
    to Li--H distances far outside the classical turning points of
    LiH. %This suggests that the conical intersection will have
    %little impact on Li--LiH collision dynamics at ultralow
    %temperatures.

\item The Li--LiH system has several possible reaction channels: an
    exchange reaction to form products identical to the reactants,
    and two insertion reactions that produce
    Li$_2$($a^3\Sigma_u^+$) and Li$_2$(X$^1\Sigma_g^+$) plus a
    ground-state hydrogen atom. The insertion reactions are highly
    endothermic, with the energy difference between the entrance
    and exit channels of the order of 12000 cm$^{-1}$ and 22500
    cm$^{-1}$ for Li$_2$(X$^1\Sigma_g^+$)+H and
    Li$_2$($a^3\Sigma_u^+$)+H, respectively.
\end{enumerate}

In a subsequent paper \cite{subs} we will analyze the dynamics of
Li--LiH collisions at ultralow temperatures, based on our best {\em ab
initio} potential. We will analyze the impact of the present
inaccuracies in the {\em ab initio} electronic
structure calculations, and discuss the prospects of sympathetic
cooling of lithium hydride by collisions with ultracold lithium atoms.

\acknowledgments{ We would like to thank Dr.~Micha{\l} Przybytek for
his invaluable technical help with the FCI calculations. We acknowledge
the financial support from the Polish Ministry of Science and Higher
Education (grant 1165/ESF/2007/03) and from the Foundation for Polish
Science (FNP) via Homing program (grant HOM/2008/10B) within EEA
Financial Mechanism. We also thank EPSRC for support under
collaborative project CoPoMol of the ESF EUROCORES Programme EuroQUAM.}

\newpage
%\bibliography{cold,filip_lih,lihli,tk_publ,ind_Prop,propag_rev,sapt2008,molpro8,ind_e1cc,ind_SAPT,peter,sexp,ind_lokal_EOM,tk_add,explicit,nowe}

\newpage
\begin{table}[h]
%\scriptsize
%\tiny
\caption{Performance of various orbital basis sets and extrapolation
schemes compared to the CCSD(T)-F12a and CCSD(T)-F12b results at the
characteristic points of the ground state potential energy surface
of Li--LiH. The notation VXZ with X=D, T, Q, and 5 denotes the
result of the orbital CCSD(T) calculations in the cc-pVXZ basis with midbond-95 set,
XY$\alpha$ with X and Y = D, T, Q, and 5, and $\alpha=2$ or 3 denote
the extrapolated result according to Eq. (\ref{extrapol}), while
F12a and F12b stand for the explicitly correlated CCSD(T) results with
the a and b approximation schemes. $\Delta_{\rm F12a}$ and $\Delta_{\rm F12b}$
are the percent error of given result with respect to the CCSD(T)-F12a
and CCSD(T)-F12b results, respectively.}
\label{tab0} \vskip 3ex
\begin{tabular}{lrrc|rrc|rrc|rrc}
\hline\hline
& \clc{$V$ (cm$^{-1}$)} & \clc{$\Delta_{\rm F12a}$} & \clc{$\Delta_{\rm F12b}$} & \clc{$V$ (cm$^{-1}$)} & \clc{$\Delta_{\rm F12a}$} & \clc{$\Delta_{\rm F12b}$}& \clc{$V$ (cm$^{-1}$)} & \clc{$\Delta_{\rm F12a}$} & \clc{$\Delta_{\rm F12b}$} & \clc{$V$ (cm$^{-1}$)} & \clc{$\Delta_{\rm F12a}$} & \clc{$\Delta_{\rm F12b}$}\\
\hline
&\multicolumn{3}{c}{Global Minimum}&\multicolumn{3}{c}{Saddle point}&\multicolumn{3}{c}{Local minimum}&\multicolumn{3}{c}{Avoided crossing}\\
\hline
VDZ   &  --8547.77  &    1.87   &   1.84  &  --1548.64  &  --0.39   & --0.61 &  --1616.54  &  --0.65   & --0.77&  --4771.50  &    4.59   &   4.65    \\
VTZ   &  --8652.38  &    0.67   &   0.64  &  --1566.01  &  --1.52   & --1.73 &  --1629.62  &  --1.46   & --1.59&  --4892.57  &    2.17   &   2.23    \\
VQZ   &  --8683.85  &    0.31   &   0.27  &  --1555.93  &  --0.86   & --1.08 &  --1618.22  &  --0.75   & --0.88&  --4955.39  &    0.91   &   0.97    \\
V5Z   &  --8698.84  &    0.14   &   0.10  &  --1551.91  &  --0.60   & --0.82 &  --1612.70  &  --0.41   & --0.53&  --4974.35  &    0.53   &   0.60    \\
\hline                                                                                                                                           
DT2   &  --8683.31  &    0.32   &   0.28  &  --1575.59  &  --2.14   & --2.36 &  --1637.22  &  --1.94   & --2.06&  --4934.30  &    1.33   &   1.40    \\
TQ2   &  --8704.14  &    0.08   &   0.04  &  --1541.85  &    0.05   & --0.17 &  --1601.47  &    0.29   &   0.17&  --4983.15  &    0.36   &   0.42    \\
Q52   &  --8714.77  &   -0.05   & --0.08  &  --1543.21  &  --0.04   & --0.25 &  --1601.76  &    0.27   &   0.15&  --4991.70  &    0.19   &   0.25    \\
\hline                                                                                                                                           
DT3   &  --8668.66  &    0.48   &   0.45  &  --1571.05  &  --1.84   & --2.06 &  --1633.62  &  --1.71   & --1.84&  --4914.53  &    1.73   &   1.79    \\
TQ3   &  --8695.37  &    0.18   &   0.14  &  --1547.94  &  --0.35   & --0.56 &  --1608.71  &  --0.16   & --0.28&  --4971.14  &    0.60   &   0.66    \\
Q53   &  --8707.24  &    0.03   &   0.01  &  --1546.78  &  --0.27   & --0.49 &  --1606.24  &  --0.01   & --0.13&  --4984.59  &    0.33   &   0.39    \\
\hline                                                                                                                                           
F12a  &  --8710.85  &    0.00   & --0.04  &  --1542.60  &    0.00   & --0.21 &  --1606.14  &    0.00   & --0.12&  --5001.10  &    0.00   &   0.06    \\
F12b  &  --8707.77  &    0.04   &   0.00  &  --1539.31  &    0.21   &   0.00 &  --1604.17  &    0.12   &   0.00&  --5004.15  &  --0.06   &   0.00    \\
\hline\hline
\end{tabular}
\end{table}

\newpage
\begin{table}[h]
%\scriptsize
%\tiny
\caption{ 
Performance of the valence-valence FCI correction $\delta V^{\rm FCI}_{\rm{v-v}}$  against exact FCI results
for the characteristic points of the LiH--Li potential (in cm$^{-1}$). Calculations were done
in the cc-pVDZ basis set. Subscript all-all refers to all electrons correlated, while v-v denotes frozen-core results. 
$\Delta$ is the percentage error of the  $\delta V^{\rm FCI}_{\rm{v-v}}$  approximation with respect to the $\delta V^{\rm{FCI}}_{\rm{all-all}}$:
 $\Delta=\frac{\delta V^{\rm FCI}_{\rm{v-v}}-\delta V^{\rm{FCI}}_{\rm{all-all}}}{|\delta V^{\rm{FCI}}_{\rm{all-all}}|}\cdot 100\%$.}

\label{tabF} \vskip 3ex
\begin{tabular}{lrrrr}
\hline\hline
                                                                      & \clc{GM}    &\clc{SP}   &\clc{LM} & \clc{AC}    \\ 
\hline                                                                
$V^{\rm CCSD(T)}_{\rm{v-v}}$                                          &  --7553.16  & --1407.50 &--1502.25&  --3785.91  \\ 
$V^{\rm FCI}_{\rm{v-v}}$                                              &  --7587.86  & --1437.27 &--1522.05&  --3806.93  \\
$\delta V^{\rm{FCI}}_{\rm{v-v}}$                                      &  --34.70    & --29.77   &--19.80  &  --21.02    \\
\hline
$V^{\rm CCSD(T)}_{\rm{all-all}}$                                      &  --7590.40  & --1415.56 &--1509.04&  --3842.01  \\
$V^{\rm FCI}_{\rm{all-all}}$                                          &  --7625.07  & --1445.32 &--1528.78&  --3863.46  \\
$\delta V^{\rm{FCI}}_{\rm{all-all}}$                                  &  --34.67    & --29.76   &--19.74  &  --21.45    \\
\hline
%$V^{\rm CCSD(T)}_{\rm{all-all}}$+$\delta V^{\rm{FCI}}_{\rm{v-v}}$    &  --7625.10  & --1445.33 &--1528.84&  --3863.30  \\ 
$\Delta [\%]$                                                         &   --0.09    & --0.04    & --0.30  &     0.76    \\
\hline\hline

\end{tabular}
\end{table}

\newpage

\begin{table}[h]
\caption{Characteristic points of the interaction potentials for the
ground state Li($^2$S) + LiH (${\rm X}\, ^1\Sigma^+$) and the first
excited state, which correlates asymptotically with Li($^2$P) + LiH
(${\rm X}\, ^1\Sigma^+$).}\label{tab1} \vskip 5ex
\begin{tabular}{lcrc}
\hline\hline
                   & $R$ [bohr] &  $\theta$ [degrees] & $V$ [cm$^{-1}$]\\
    \hline
    \multicolumn{4}{c}{Ground state}\\
    \hline
    Global minimum & 4.40     &  46.5$^\circ$                & --8743      \\
    Local  minimum & 6.56     & 180.0$^\circ$                & --1623      \\
    Saddle point   & 6.28     & 136.0$^\circ$                & --1565      \\
    \hline
    \multicolumn{4}{c}{Excited state}\\
    \hline
    Global minimum & 5.66     &   0.0$^\circ$                & --4743      \\
\hline\hline
\end{tabular}
\end{table}

\newpage
\begin{table}[h]
\caption{Long-range coefficients (in atomic units) for Li--LiH, from
perturbation theory. Numbers in parentheses indicate powers of 10.
\label{tab2}} \vskip 5ex
\begin{tabular}{lrrrrrr}
\hline\hline
$ l ~ \rightarrow$      & \clc{0}  &  \clc{1} &  \clc{2}  &  \clc{3}  &  \clc{4}  \\
\hline
$C_6^{l}$               &   1247.8 &            &  869.7  &           &           \\
$C_7^{l}$               &          &   8304.2   &         &  2902.3   &           \\
$C_8^{l}$               &   4.83(4)&            &  4.24(4)&           &   8923.0  \\
$C_9^{l}$               &          &   3.94(5)  &         &  6.42(4)  &           \\
$C_{10}^{l}$            &   2.03(6)&            &  2.02(6)&           &   2.44(5) \\
\hline\hline
\end{tabular}
\end{table}

% \newpage
% \begin{table}[h]
% \caption{Components of the interaction energy from SAPT calculations at
% the global minimum (GM), saddle point (SP), and local minimum (LM).
% Energies are given in cm$^{-1}$. \label{tab3}} \vskip 5ex
% \begin{tabular}{lrrr}
% \hline\hline
%                             & \clc{GM} & \clc{SP} & \clc{LM} \\
% \hline
% $E^{(1)}_{\rm elst}$         & --14442 &   251   &     327  \\
% $E^{(2)}_{\rm ind}$          & --20539 &--3561   &  --3393  \\
% $E^{(2)}_{\rm disp}$         &  --6095 & --447   &   --288  \\
% $E^{(1)}_{\rm exch}$         &   20189 &   986   &     751  \\
% $E^{(2)}_{\rm exch-ind}$     &    9955 &  1535   &    1332  \\
% $E^{(2)}_{\rm exch-disp}$    &     230 &    74   &      42  \\
% $E_{\rm exch}$               &   30374 &  2595   &    2125  \\
% \hline\hline
% \end{tabular}
% \end{table}

\newpage

%\begin{tabular}{ccc}
%{\includegraphics[scale=0.30,angle=0]{GlobalMinimum.eps}} &\hspace*{1.0cm}
%{\includegraphics[scale=0.30,angle=0]{SaddlePoint.eps}}  &\hspace*{1.0cm}
%{\includegraphics[scale=0.30,angle=0]{LocalMinimum.eps}}

\newpage

\newpage

\newpage
\newpage
\newpage
\newpage
\clearpage
\clearpage

\end{document}